%% file: main.tex
\documentclass[10pt,conference]{IEEEtran}

\usepackage{booktabs} % For formal tables
\usepackage{amsmath,amsfonts,mathtools,bbm,amssymb}
\usepackage{graphicx}
\usepackage{textcomp}
\usepackage{comment}
\usepackage{soul}
\usepackage{url}
\usepackage{enumitem}
\usepackage{footnote}
\usepackage{xcolor}
\usepackage[binary-units]{siunitx}
\usepackage{todonotes}
\usepackage[linesnumbered,ruled,vlined,noend]{algorithm2e}
\usepackage{pgfplots,multirow}
\usepackage{xspace}
\usepackage{subcaption}
\usepackage[hidelinks]{hyperref}
\usepackage[maxfloats=256]{morefloats}
\pgfplotsset{compat=1.11}

\DeclareRobustCommand{\IEEEauthorrefmark}[1]{\smash{\textsuperscript{\footnotesize #1}}}

\SetCommentSty{mycommfont}

\newif\iffinal
\finaltrue 

\iffinal
    \newcommand{\remove}[1]{}
    \newcommand{\removeeq}[2]{}
    \newcommand{\revise}[1]{#1}
\else
    \newcommand{\remove}[1]{{\color{red}\sout{#1}}}
    \newcommand{\removeeq}[2]{{\color{red}\begin{equation}\text{\sout{\ensuremath{#2}\tag{\sout{#1}}}}\end{equation}}}
    \newcommand{\revise}[1]{{\color{blue}#1}}
\fi

\input{macros}

\begin{document}
\setlength{\abovedisplayskip}{2mm}
\setlength{\belowdisplayskip}{2mm}
\setlength{\belowdisplayshortskip}{2pt}
\setlength{\abovedisplayshortskip}{2pt}

\title{Co-Optimizing Cache Partitioning and Multi-Core Task Scheduling: Exploit Cache Sensitivity or Not?}

\author{
\IEEEauthorblockN{Binqi Sun\IEEEauthorrefmark{1},
Debayan Roy\IEEEauthorrefmark{1},
Tomasz Kloda\IEEEauthorrefmark{2},
Andrea Bastoni\IEEEauthorrefmark{1}, 
Rodolfo Pellizzoni\IEEEauthorrefmark{3} and
Marco Caccamo\IEEEauthorrefmark{1}}
\IEEEauthorblockA{\IEEEauthorrefmark{1}Technical University of Munich, Germany}
\IEEEauthorblockA{\IEEEauthorrefmark{2}LAAS-CNRS, Université de Toulouse, INSA, Toulouse, France}
\IEEEauthorblockA{\IEEEauthorrefmark{3}University of Waterloo, Canada
\\Email: \{binqi.sun, debayan.roy, andrea.bastoni, mcaccamo\}@tum.de, tkloda@laas.fr, rpellizz@uwaterloo.edu}
}

\maketitle

\begin{abstract}
Cache partitioning techniques have been successfully adopted to \emph{mitigate interference} among concurrently executing real-time tasks on multi-core processors. 
Considering that the execution time of a cache-sensitive task strongly depends on the cache available for it to use, \emph{co-optimizing} cache partitioning and task allocation improves the system's \emph{schedulability}.  
In this paper, we propose a \emph{hybrid multi-layer} design space exploration technique to solve
this \emph{multi-resource} management problem.
We explore the \emph{interplay} between cache partitioning and schedulability by systematically \emph{interleaving} three optimization layers, \textit{viz.},
(i)~in the outer layer, we perform a \emph{breadth-first search} combined with proactive \emph{pruning} for cache partitioning; (ii)~in the middle layer, we exploit a \emph{first-fit} heuristic for allocating tasks to cores; and (iii)~in the inner layer, we use the well-known recurrence relation for the schedulability analysis of \emph{non-preemptive fixed-priority (NP-FP) tasks} in a uniprocessor setting.
Although our focus is on NP-FP scheduling, we evaluate the flexibility of our framework in supporting different scheduling policies (NP-EDF, P-EDF) by plugging in appropriate analysis methods in the inner layer.
Experiments show that, compared to the state-of-the-art techniques, the proposed framework can \emph{improve the real-time schedulability of NP-FP task sets by an average of 15.2\%}
with a maximum improvement of 233.6\% (when tasks are highly cache-sensitive) and a minimum of 1.6\% (when cache sensitivity is low).
For such task sets, we found that \emph{clustering} similar-period (or \emph{mutually compatible}) tasks 
often leads to higher schedulability (on average 7.6\%)
than clustering by \emph{cache sensitivity}.
In our evaluation, the framework also achieves good results for preemptive and dynamic-priority scheduling policies.
\end{abstract}

\input{sections/intro}

\input{sections/relatedworks}

\input{sections/problem}

\input{sections/algorithms}

\input{sections/algex}
\input{sections/results}

\input{sections/conclusion}

\section*{Acknowledgement}
Marco Caccamo was supported by an Alexander von Humboldt Professorship endowed by the German Federal Ministry of Education and Research.

\bibliographystyle{IEEEtran}
\bibliography{bibliography}

\input{appendix}

\end{document}

%% file: macros.tex
\definecolorseries{test}{rgb}{grad}[rgb]{.95,.55,.55}{11,11,17}
\resetcolorseries[10]{test}
\newcommand{\addtodoeditor}[1]{%
    \colorlet{#1}{test!!+!50}
    \expandafter\newcommand\csname#1\endcsname [1]{%
        \todo[color=#1,size=\tiny]{\sffamily\textbf{\uppercase{#1}:}
    ##1}\xspace%
    }
    \expandafter\newcommand\csname#1i\endcsname [1]{%
        \todo[inline, color=#1]{\sffamily\textbf{\uppercase{#1}:} ##1}\xspace%
    }
}

\addtodoeditor{bs}
\addtodoeditor{ab}
\addtodoeditor{dr}
\addtodoeditor{tk}
\addtodoeditor{rp}

\newcommand{\eg}{{\it e.g.}\xspace}
\newcommand{\ie}{{\it i.e.}\xspace}

\let\oldnl\nl% Store \nl in \oldnl
\newcommand{\nonl}{\renewcommand{\nl}{\let\nl\oldnl}}% Remove line number for one line

%% file: sections/intro.tex
\section{Introduction}\label{sec:intro}
\begin{figure}[t]
    \centering
        \includegraphics[width=0.9\columnwidth]{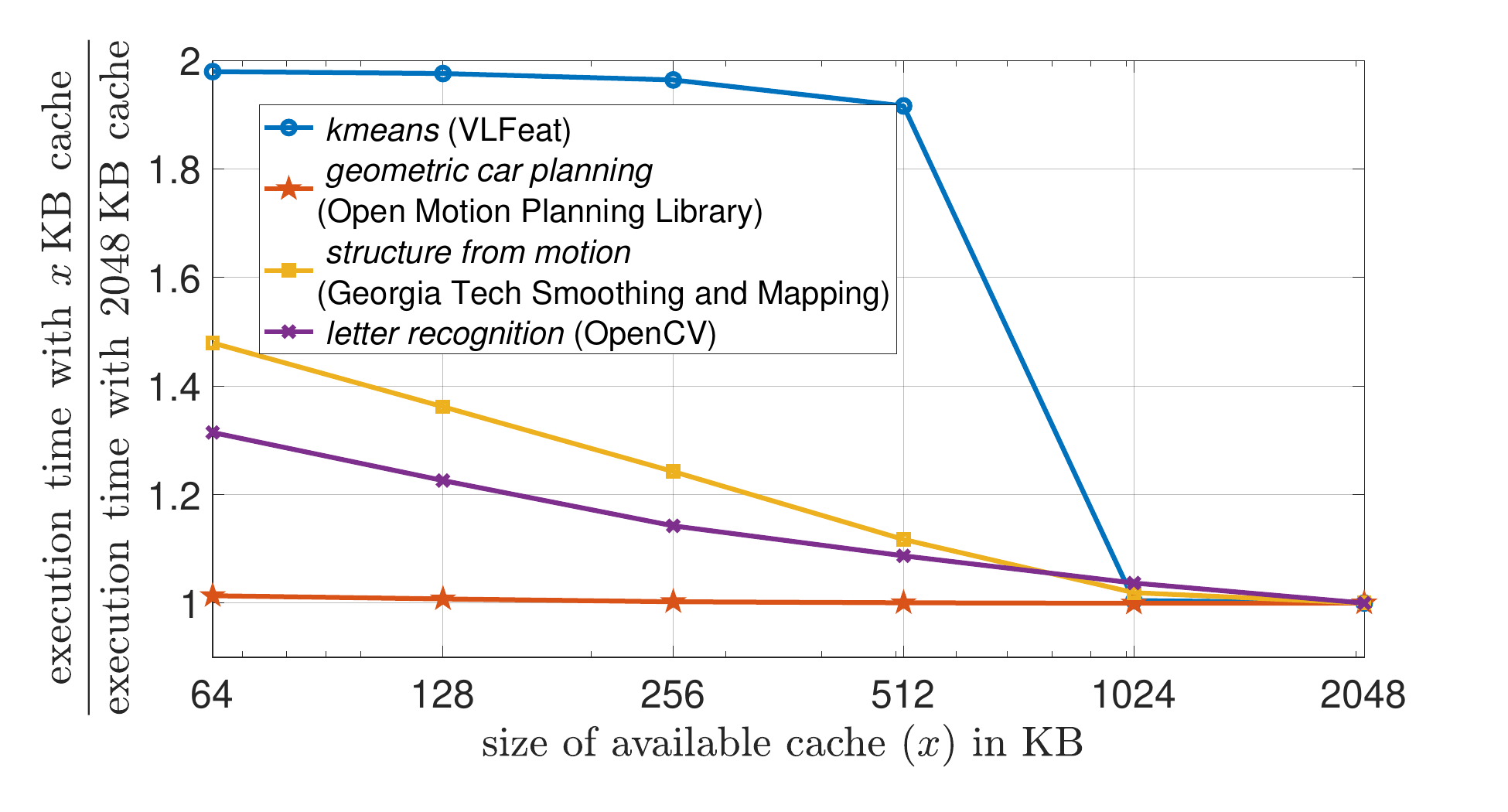}
        \vspace{-2mm}
    \caption{Benchmark's execution slowdown with $x$~KB cache compared to full (2048~KB) cache.}
    \vspace{-6mm}
\label{fig:CaseStudy}
\end{figure}

Nowadays, heterogeneous multiprocessor system-on-a-chip (MPSoC) platforms are routinely used for all those workloads that require performance, real-time capabilities, and limited size and power consumption. 
These workloads include, \eg, applications found in autonomous driving, intelligent robotics, and unmanned aerial vehicles domains.
Towards guaranteeing real-time performance, 
these platforms pose an unprecedented challenge to the management of the memory hierarchy.
With a focus on the core complex
of such MPSoCs,
sharing caches among cores prevents analyzing tasks in isolation, thus
complicating an accurate estimation of the tasks' worst-case execution times (WCETs).
Unsurprisingly, therefore, to mitigate this problem, both software-based ~\cite{SM:08, %GF:13, 
WHKA:13, MDBCCP:13, KSMCV:19} and hardware-based~\cite{XPCLLLL:19, SSRM:12} cache partitioning techniques have been exploited.
Although effective, cache partitioning limits the amount of cache available to (groups of) real-time tasks. 
Therefore, the impact of cache partitioning on the WCET can be non-negligible for a \emph{cache-sensitive} workload.
This effect is illustrated in Figure~\ref{fig:CaseStudy}, which reports the \emph{slowdown} due to reduced cache availability of four benchmark applications (more details in Section~\ref{sec:results}).
For example, \emph{kmeans} is almost two times slower when it runs with a cache partition smaller than 256~KB instead of 1024~KB.

\smallskip
\noindent\textbf{Problem setting:}
This paper studies the \emph{integrated problem} of (1)~assigning real-time tasks to cores and (2)~reserving cache for tasks running on each core.
The goal is to achieve a solution where the tasks are \emph{schedulable}, \ie, each task meets its real-time requirements, \eg, deadline. 
The main focus of the proposed optimization strategies is
on \emph{non-preemptive fixed-priority} (NP-FP) scheduling. We further assume that tasks are \emph{statically assigned} to cores.
Partitioned schedulers have simpler implementations
and generally lower overheads~\cite{BBA:10},
and non-preemption naturally separates
computation from data management phases (\eg,~\cite{TMWPC:19}).
Also, we assess the flexibility of our framework to support other scheduling policies such as preemptive and non-preemptive \textit{earliest deadline first} (EDF).

\smallskip
\noindent
\textbf{Proposed framework:}
Given the \emph{interdependencies} of the three sub-problems, viz., task allocation, cache partitioning, and schedulability analysis, this paper studies an \emph{integrated} solution to improve the likelihood of establishing system schedulability.  
In particular, we propose a \emph{nested multi-layer, hybrid} optimization framework to \emph{explore the interplay} between the sub-problems.
In this framework, (i)~the outer layer partitions the shared cache, (ii)~the middle layer allocates tasks, and (iii)~the inner layer performs the schedulability analysis. 

We perform a \emph{polynomial-time breadth-first search} in the outer layer
using
a heuristic to \emph{proactively prune} the search tree to prevent its exponential growth (Section~\ref{sec:outer}).
The outer layer chooses a cache partition size for a core and invokes the middle layer to allocate tasks to the core from the remaining ones.
We develop two strategies for task allocation. 
While one tries to cluster tasks that are \emph{compatible} for co-scheduling, the other one co-allocates tasks with similar \emph{cache sensitivity potentials} (Section~\ref{sec:inner}).
The schedulability of the task set allocated by the middle layer is checked by the inner layer using the exact method reported in~\cite{Davis:2007} for NP-FP scheduling.

Although the selection of tasks is optimized for NP-FP scheduling, the multi-layer framework can be easily adapted to other scheduling policies by plugging in an appropriate schedulability test in the inner layer.
To demonstrate this, we show experimental results under
\textit{preemptive EDF} (P-EDF) and \textit{non-preemptive EDF} (NP-EDF) scheduling by using the tests adopted by~\cite{XPCLLLL:19} and~\cite{Paolieri:2011, Berna:2012}, without any modifications to the outer and middle layers (Section~\ref{sec:results_edf}).

\smallskip
\noindent\textbf{Contributions:} This paper has the following contributions:
\begin{itemize}[leftmargin=*]
    \item A generic multi-layer, hybrid optimization framework
    is proposed to solve the joint problem of cache partitioning and partitioned scheduling of real-time tasks.
    In particular, we systematically extend
    the first-fit heuristic for task allocation with an outer layer employing an intelligent breadth-first search for cache partitioning.
    \item To the best of our knowledge, this paper shows for the first time that the characteristics of task sets (including task periods and cache sensitivities)
    guide the choice of heuristics to solve the aforementioned problem. 
    \item A metric is introduced to evaluate the cache sensitivity potential of a task that assists in task allocation. This metric captures the maximum possible reduction in the utilization of a task if more cache can be offered.
    \item The framework is extensively evaluated and compared with multiple state-of-the-art techniques~\cite{XPCLLLL:19,Paolieri:2011,Berna:2012},
    using both benchmark-derived and synthetic cache slowdown profiles.
    Results for NP-FP scheduling indicate that the performance of the framework depends on the cache-sensitivity of workloads with
    a schedulability improvement of up to 14.5\% for tasks with low cache sensitivity and of up to 233.6\% for highly cache-sensitive tasks, with an average improvement of 15.2\%.
    NP-FP experiments also show that focusing on compatibility leads to better results (by 7.6\% on average) than cache sensitivity. %while clustering tasks in the middle layer.
    For P-EDF scheduling, the framework improves the schedulability by 8.7\% on average compared to the approach in~\cite{XPCLLLL:19}, while for NP-EDF scheduling, the average improvement is 19.2\% compared to the techniques in~\cite{Paolieri:2011, Berna:2012}.
    \end{itemize}

\smallskip
\noindent
\textbf{Paper organization:}
We present relevant previous works in Section~\ref{sec:related}.
We define the problem in Section~\ref{sec:prob}. We describe our proposed framework and heuristics in Section~\ref{sec:algo}. In Section~\ref{sec:algex}, using illustrative examples, we show that neither of the proposed heuristics dominates the other. Experimental results and interesting trends are discussed in Section~\ref{sec:results}. Section~\ref{sec:concl} provides concluding remarks.

%% file: sections/relatedworks.tex
\section{Related Works}\label{sec:related}

\smallskip
\noindent
\textbf{Preemptive task-cache co-allocation:}
Full exploitation of multiprocessor platforms can be achieved only if the allocation of tasks and memory (\eg, cache) resources is performed jointly.
\emph{Tunable} WCETs~\cite{Yoon:2011} can be elastically adjusted to take into account shared resource allocation and arbitration methods. For this model,
mixed-integer linear programming (MILP) has been used to partition tasks, cache, and bandwidth, minimizing the overall system utilization.
In~\cite{Suzuki13}, tasks, cache, and bandwidth co-allocation problem is solved using a MILP formulation and
a knapsack-algorithm-based heuristic. \cite{Suzuki13} considers P-EDF scheduling while allocating dedicated cache partitions to tasks.
Likewise,~\cite{Sarkar:2015} considers recurrent tasks scheduled under P-EDF and proposes a different heuristic for task and cache allocation on a multiprocessor.
Under partitioned P-EDF, the dependence of execution time on the number of available cache partitions has been studied in \cite{XPCLLLL:19, CWKA:15, KWCFAS:17}.
\cite{XPCLLLL:19} uses k-means clustering and a first-fit heuristic to partition shared caches and allocate tasks on a multiprocessor based on cache sensitivities of tasks. 
\cite{XPCLLLL:19} also compares the obtainable schedulability with the strategy adopted by~\cite{CWKA:15} (and later also used in~\cite{KWCFAS:17, bakita2021simultaneous}).
The above works mainly consider setups for which exact utilization bound tests are available, which is not true for our setup.
Nevertheless, in Section~\ref{sec:results}, we compare our framework with~\cite{XPCLLLL:19} with appropriate adaptations for non-preemptive and preemptive scheduling.

\smallskip
\noindent
\textbf{Non-preemptive task-cache co-allocation:}
For non-preemptive scheduling, the problem becomes more complex as there are no efficient utilization bounds to check the schedulability in polynomial time.
The multi-resource allocation problem for time-triggered non-preemptive scheduling naturally fits into integer linear programming (ILP) formulation~\cite{Chen14}.
\emph{Interference Aware Allocation Algorithm} (IA$^3$)~\cite{Paolieri:2011} and \emph{Period Driven Task and Cache Partitioning
Algorithm} (PDPA)~\cite{Berna:2012} are proposed for NP-EDF. 
These works are the closest to ours. 
While, qualitatively, we search the design space more thoroughly than them using a carefully designed pruning criterion and a cache sensitivity potential metric.
In Section~\ref{sec:results}, we quantitatively compare our proposed approach to them (for both NP-FP and NP-EDF).
Also, we empirically show that the characteristics of task sets guide the selection of the heuristics to solve the problem.

\smallskip
\noindent
\textbf{Cache-related preemption delays (CRPD):}
In multitasking systems, partitioning shared caches and allowing each task to run using specific partitions can reduce CRPD, but it may increase cache misses.
\cite{ADM:12} shows how to systematically compute CRPD.
Optimally exploring the trade-off between CRPD and cache misses, even for a single-core, is NP-hard~\cite{Bui:2008}.
Several algorithms~\cite{Bui:2008,Altmeyer:2016,sun2023minimizing} have been proposed in this context to optimize task set utilization and cache usage.
In the same vein, \cite{Kim_cache} considers sharing cache partitions among tasks running preemptively on the same core while allowing cache isolation between cores on a multiprocessor. The task and cache allocations are performed using best-fit decreasing bin-packing. The approach is also extended for multi-core virtualization~\cite{Kim16:EMSOFT}.
Contrary to these works, we consider that only NP-FP tasks running on a core can share cache partitions and, hence, they do not experience CRPD.

\smallskip
\noindent
\textbf{Other approaches:}
Instead of partitioning shared caches to cores, \cite{xiao2022cache} formulates an ILP to upper bound the inter-core cache interference and proposes a task partitioning algorithm based on the interference upper bound.
Dynamic resource allocation has also been explored, \eg, \cite{gifford2021dna} adapts the resource allocation based on program phases, while \cite{Kwon:2021, gifford2022multi} dynamically allocate resource at mode transitions.

%% file: sections/problem.tex
\section{Problem Description}\label{sec:prob}
In this section, we describe the design problem under study. In particular, we want to (i)~determine an allocation of software tasks to cores and (ii)~identify the maximum portion of cache each task can use so that (iii) the tasks meet their respective deadlines. 
Table~\ref{tab:notation} summarizes the most important symbols used in the following sections.

\input{tables/notation_simplified}

\subsection{Task allocation}
We consider a set of \revise{$n_\tau$} software tasks, denoted by $\mathcal{T} = \{\tau_1,\tau_2,\cdots,\tau_{n_\tau}\}$.
These tasks need to run on $n_c$ processing cores. 
We denote the set of processors by $\mathcal{C} = \{C_1,C_2,\cdots,C_{n_c}\}$.
For this setting, we need to determine \emph{how the tasks should be allocated to the cores}, which is the first part of the problem at hand. 
Hence, we determine $T_1,T_2,\cdots,T_{n_c}$, where $T_j$ is a set of tasks that will run on the core $C_j$.
We study \emph{partitioned} multi-core scheduling, \ie, a task $\tau_i \in T_j$ allocated to a core $C_j$ will always be executed by $C_j$.
It means that the task allocation is static.
In mathematical terms, $\{T_1,T_2,\cdots,T_{n_c}\}$ is a partition of set $\tau$.

\subsection{Cache partitioning}\label{sec:cachepartitioning}

We assume that tasks execute on cores that share a cache of size $M$ (this model is common in current MPSoCs, \eg,~\cite{zcu-102, nvidia-agx-xavier}). 
To prevent memory interference, we consider dividing the shared cache into $n_p$ partitions of equal size. The tasks assigned to a core can only use a certain number of partitions.
Our approach applies to \emph{any} cache-partitioning technique, either with hardware-support (\eg, \cite{intel-cat}, \cite{arm-mpam}), or purely implemented in software (\eg,~\cite{MDBCCP:13, KSMCV:19}).

We reserve only a certain number of cache partitions for each core and let the tasks running on a core use only the assigned partitions. 
The amount of cache that the tasks running on a core optimally need depends on how their execution times vary with available cache and their real-time requirements.
Hence, we need to determine \emph{the appropriate number of cache partitions $\mu_j$ that should be made available to the tasks (in $T_j$) mapped on a core $C_j$}, which is the second part of the problem under consideration. 
In the meantime, we need to respect the constraint given by 
$\sum_{j=1}^{n_c} \mu_j \leq n_p$,
\ie, the total number of partitions in use cannot be more than $n_p$.

\subsection{Task specification} 
We consider that each non-preemptive task $\tau_i \in \mathcal{T}$ is dispatched sporadically, respecting a minimum time $p_i$ between two consecutive dispatches. In this paper, we also refer to $p_i$ as the period of the task $\tau_i$.
We study the case where the deadline of a task $\tau_i$ is exactly equal to its period $p_i$. 
The algorithms presented in this paper are valid or can be trivially extended for other deadline constraints as well.
The execution time $e_i$ of a task $\tau_i$ depends on the number of cache partitions it can use.
Hence, we can write $e_i = \mathcal{E}_i(\mu)$, where $\mathcal{E}_i(\cdot)$ is a discrete function of the available number of cache partitions
$\mu \in \{1,2,\cdots, n_p\}$. We can write the range of $\mathcal{E}_i(\cdot)$ as an ordered set $(\epsilon_{i,1},\epsilon_{i,2},\cdots,\epsilon_{i,n_p})$, where $\epsilon_{i,\mu}$ is the execution time when $\tau_i$ uses $\mu$ cache partitions. 
Note that we assume a task will use at least one cache partition.
In summary,
we specify a task $\tau_i$ using its period (or minimum inter-arrival time) $p_i$ and the function $\mathcal{E}_i(\cdot)$ capturing the variation of its execution time with the available number of cache partitions.
We assume that $p_i$ and $\mathcal{E}_i(\cdot)$ can be computed a priori for $\tau_i$, and $\mathcal{E}_i(\cdot)$ includes scheduling overheads and all other sources of interference (\emph{e.g.,~inter-core interference}).
We note that different solutions can be used alongside to mitigate the timing interference in modern multi-core platforms (\emph{e.g.,}~bank partitioning~\cite{Pan:2018,Cheng:2017,Yun:2014}, software bandwidth regulators~\cite{yun2013memguard,Kritikakou:2014} or segmented execution models~\cite{Pellizzoni:2011,PremTech,Durrieu:2014,Melani,Arora}).
\revise{In this work, we assume that $\mathcal{E}_i(\cdot)$ is determined while each core gets equal memory bandwidth using, \eg, MemGuard~\cite{yun2013memguard}.
}

\subsection{Schedulability analysis}\label{sec:schedanalysis} 
We need to \emph{verify the schedulability of a set of tasks $T_j$} that is allocated to a core $C_j$ and uses $\mu_j$ cache partitions to run, which is the third part of the problem under study. 
For a task $\tau_i \in T_j$, we can write $e_i = \epsilon_{i,\mu_j}$. 
We consider that a task cannot be preempted during execution. 
Besides, each task has a fixed priority according to the rate monotonic scheduling policy. 
A task $\tau_i$ has a higher priority than a task $\tau_{i'}$ (where $\tau_i,\tau_{i'} \in T_j$) if $p_i < p_{i'}$.
When $p_i = p_{i'}$, we assume that $\tau_i$ has a higher priority than $\tau_{i'}$ if $e_i > e_{i'}$. 
In all other cases, $\tau_i$ has a lower priority than $\tau_{i'}$. 
Here, no two tasks mapped on the same core have the same priority.

We compute the worst-case response time of a task under the NP-FP scheduling policy using the technique outlined in~\cite{Davis:2007}.
First, we determine the busy period $t_i$ of a task $\tau_i \in T_j$ using the following recurrence relation: 
\begin{equation}\label{eq:busyperiod}
    t_i^{k+1} = B_{i,j} + \sum_{\tau_{i'} \in HEP_{i,j}} \left\lceil \frac{t_i^k}{p_{i'}} \right\rceil e_{i'}
\end{equation}
where, $B_{i,j} = \max_{\tau_{i'} \in LP_{i,j}} e_{i'}$.
Here, (i)~$HEP_{i,j}\subseteq T_j$, $LP_{i,j} \subset T_j$ where $HEP_{i,j}$  and $LP_{i,j}$, respectively, comprise the tasks that have higher or equal and lower priorities than $\tau_i$; and (ii)~$B_{i,j}$ is the maximum time for which the task $\tau_i$ can be blocked by a lower priority task.
To solve Equation~\ref{eq:busyperiod}, we start with $t_i^0 = e_i$ and continue until we get $t_i^{k+1} = t_i^{k}$.
Here, the recurrence relation is guaranteed to converge if $\sum_{\tau_{i'} \in HEP_{i,j}} \frac{e_{i'}}{p_{i'}} < 1$.
Further, we calculate the number of instances $Q_i$ of $\tau_i$ that execute in its busy period $t_i$ as follows:
\begin{equation}
    Q_i = \left\lceil \frac{t_i}{p_i} \right\rceil.
\end{equation}
We compute the response time for each of the $Q_i$ instances. We can calculate the longest time $w_i(q)$ between the start of the busy period and the start of the execution of the $q$-th instance ($1\leq q \leq Q_i$) of $\tau_i$ using the following recurrence relation:
\begin{equation}\label{eq:startexec}
    w_i^{k+1}(q) = B_{i,j} + (q-1) \cdot e_i + \sum_{\tau_{i'} \in HP_{i,j}} \left\lceil \frac{w_i^{k}(q) + \delta}{p_{i'}} \right\rceil e_{i'}.
\end{equation}
Here, (i)~$HP_{i,j}\subset T_j$ comprises the tasks that have higher priorities than $\tau_i$; and (ii)~$\delta > 0$ is a very small number.
To solve the recurrence relation in Equation~\ref{eq:startexec}, we start with $w_i^0(q) = B_{i,j} + (q-1)\cdot e_i$ and stop when $w_i^{k+1}(q) = w_i^{k}(q)$.
The response time of the $q$-th instance of $\tau_i$ is given as follows:
\begin{equation}
    R_i(q) = w_i(q) - (q-1) \cdot p_i + e_i.
\end{equation}
The worst-case response time $R_i^{wc}$ is computed as follows:
\begin{equation}
    R_i^{wc} = \max_{1\leq q \leq Q_i} R_i(q).
\end{equation}
An implicit-deadline task $\tau_i$ meets its deadline if and only if
\begin{equation}\label{eq:deadline}
    R_i^{wc} \leq p_i.
\end{equation}
A task set is schedulable if and only if Equation~\ref{eq:deadline} holds for each task in the task set.

%% file: tables/notation_simplified.tex
\begin{table}[t]
  \begin{center}
    \caption{List of symbols.}
    \label{tab:notation}
    \scriptsize
    \renewcommand{\arraystretch}{1.1}
    \begin{tabular}{|c|l|}
      \hline
      \textbf{Symbols} & \textbf{Description} \\ \hline \hline
      $\mathcal{T}$ & set of $n$ tasks $\mathcal{T}=\{\tau_1,\tau_2,\dots,\tau_{n_\tau}\}$; \\ \hline
      $\mathcal{C}$ & set of processors $\mathcal{C}=\{C_1,C_2,\dots,C_{n_c}\}$; \\ \hline
      $n_p$ & total number of cache partitions; \\ \hline
      $p_i$ & period of task $\tau_i$; \\ \hline
      $\epsilon_{i,\mu}$ & execution time of $\tau_i$ with $\mu$ cache partitions; \\ \hline
      $T_j$ & set of tasks allocated to core $C_j$; \\ \hline
      $\mu_j$ & number of cache partitions allocated to core $C_j$; \\ \hline
      $\gamma_i$ & cache sensitivity potential of task $\tau_i$; \\ \hline
      $\widehat{u}_i$ & base utilization of task $\tau_i$, $\widehat{u}_i=\epsilon_{i,n_p}/p_i$;   \\ \hline
      $\omega_{init}$ & an empty solution; \\ \hline
      $\omega,\omega',\omega^+$ & a (new) partial solution; \\ \hline
      $\Omega^*$ & set of new partial solutions; \\ \hline
      $\Omega_x$ & set of partial solutions at search depth $x$; \\ \hline
      $\mathcal{T}^{\>*}$ & set of remaining tasks to be allocated; \\ \hline
      $\overline{\mathcal{T}^{\>*}}$ & sorted list of remaining tasks to be allocated; \\ \hline
      $U_R$ & scheduling demand of remaining tasks to be allocated; \\ \hline
    \end{tabular}
  \vspace{-6mm}
  \end{center}
\end{table}

%% file: sections/algorithms.tex
\vspace{-1mm}
\section{Multi-Layer Hybrid Optimization}\label{sec:algo}
As described in Section~\ref{sec:prob}, the problem under study comprises three interdependent parts.
In this paper, we propose an integrated solution with three nested layers.
The outer layer partitions the shared cache, the middle layer allocates tasks, and the inner layer performs schedulability analysis.
Algorithm~\ref{alg:inner} outlines the middle and the inner layers, while Algorithm~\ref{alg:outer} captures the outer layer.

\subsection{Algorithm~\ref{alg:inner}: Middle and inner optimization layers}\label{sec:inner} 

We provide a set of remaining tasks $\mathcal{T}^{\>*}$ and their timing attributes as input to Algorithm~\ref{alg:inner}.
The period $p_i$ and the least-possible execution time $\epsilon_{i,n_p}$ (\ie, when the whole cache is available) are known a priori for a task $\tau_i \in \mathcal{T}^{\>*}$. 
Further, in the inner two layers, we deal with a fixed number of cache partitions $\mu$ as chosen by the outer layer (more details on the outer layer in Sec.~\ref{sec:outer}).
Corresponding to $\mu$, the execution time $e_i$ of a task $\tau_i \in \mathcal{T}^{\>*}$ also gets a fixed value as $e_i = \epsilon_{i,\mu}$.

\input{Algorithms/algo_1}

The inner two layers select a set of tasks $T \subseteq \mathcal{T}^{\>*}$ that can be allocated to a core with a given amount of cache $\mu$ without violating schedulability constraints.
The goal is to increase the likelihood of scheduling the tasks in $\mathcal{T}^{\>*}\setminus T$ on the remaining cores with the rest of the cache.
There are two major challenges here. \revise{First, the number of possible selections is exponential with respect to the number of tasks in $\mathcal{T}^{\>*}$ and, hence, it is \emph{computationally hard} to go through each of them.}
Second, it is non-trivial to identify a metric to optimize.

\revise{
We apply a ``\textit{first-sort-then-pack}'' heuristic to tackle the first challenge. First, the tasks are sorted according to some criterion in Algorithm~\ref{alg:inner} line~2. Then, in lines~3~-~5, we iterate through the tasks in the sorted list $\overline{\mathcal{T}^{\>*}}$. In each iteration, we take a task and check if we can add it to the list of selected tasks $T$ without jeopardizing the schedulability. 
Note that in the inner layer (\ie, line~4), we check the schedulability of the tasks in $T$ together with the new task $\tau_i$ using the exact analysis from~\cite{Davis:2007} as outlined in Section~\ref{sec:schedanalysis}. 

To solve the second challenge, we identify two deciding factors (\ie, mutual compatibility and cache-sensitivity potential) that could affect the likelihood of scheduling remaining tasks with remaining resources. Furthermore, we utilize these two factors to propose two task sorting criteria (\ie, \textit{COMP} and \textit{CASE}) to be used in Algorithm~\ref{alg:inner}~line~2.
}

\smallskip
\noindent
\revise{\textbf{\textit{Deciding Factor~1: Cache-sensitivity potential.}}}
To elaborate on the first deciding factor, let us first define the \emph{base utilization} $\widehat{u}_i=\frac{\epsilon_{i,n_p}}{p_i}$ of a task as the ratio of its execution time and period when the whole cache is available for it to use. 
Now, the \emph{scheduling demand} $U_R$ of the tasks remaining $\mathcal{T}^{\>*}\setminus T$ after Algorithm~\ref{alg:inner} has operated on $\mathcal{T}^{\>*}$ can be defined as the sum of the base utilizations of these tasks, \ie, 
\begin{equation}\label{eq:scheddemand}
    U_R = \sum_{\tau_i \in \mathcal{T}^{\>*}\setminus T} \widehat{u}_i.
\end{equation}
Intuitively, we would like to minimize $U_R$. 
Note that the execution time $e_i$ of a task $\tau_i \in \mathcal{T}^{\>*}$ is fixed in the inner layers.
Accordingly, the utilization $u_i = \frac{e_i}{p_i}$ of the task materializes once it is selected by Algorithm 1 to be allocated to a core.
We define \emph{cache sensitivity potential} $\gamma_i$ as the difference between the utilization of
a task $\tau_i$ if it is selected by Algorithm~\ref{alg:inner} and its base utilization $\widehat{u}_i$, \ie, %$\gamma_i = \frac{e_i}{p_i} - \widehat{u}_i$.
\begin{equation}
    \gamma_i = u_i - \widehat{u}_i.
\end{equation}
The lower the value of $\gamma_i$, the lower the cache sensitivity potential of $\tau_i$.
Intuitively, the ``potential'' expresses the possibility of considerably reducing $e_i$ by providing more cache partitions.
Hence, postponing allocating a task with a lower potential might not enhance the schedulability because we cannot reduce its utilization significantly.
From another perspective, let us consider two tasks $\tau_1$ and $\tau_2$ with $\widehat{u}_1= 0.15$, $\widehat{u}_2=0.2$, $\gamma_1=0.06$, and $\gamma_2=0.01$. Here, each task will use \SI{21}{\percent} of the processor time 
and let us assume that we can allocate only one of them. When we allocate $\tau_1$, the remaining scheduling demand $U_R$ will be higher than that obtained by allocating $\tau_2$. Hence, to follow the intuition of minimizing $U_R$, we would like to add tasks with lower values of cache sensitivity potentials.

\smallskip
\noindent
\revise{\textbf{\emph{Sorting Criterion~1: CASE}.}}
Based on the above discussion, we propose to sort the tasks in a non-decreasing order of their cache sensitivity potentials. We term our multi-layer optimization with this sorting criterion as \emph{CASE}.

\smallskip
\noindent
\revise{\textbf{\textit{Deciding Factor~2: Mutual compatibility.}}}
\emph{Unlike in P-EDF scheduling},
utilization is not the only deciding factor for non-preemptive tasks. 
For example, let us consider two scenarios: 
(i)~two remaining tasks have the same period of 80 time units and base utilizations of 0.15 and 0.2, respectively; 
(ii)~two remaining tasks have periods of 10 and 80 time units, respectively, and base utilizations of 0.1 and 0.2, respectively. 
In scenario~(ii), $U_R=0.3$, while in scenario~(i), $U_R = 0.35$. 
Intuitively, we would like to end up in scenario~(ii). 
However, in this scenario, the two tasks are not schedulable on one core because the lower priority task might block the higher priority task for a time (at least 16 time units) greater than its deadline (10 time units).
In the rest of the discussion, we term two tasks to be
\emph{mutually incompatible} if they are not schedulable together on a core despite their utilizations adding up to less than or equal to 1.
When the number of remaining cores is more than one and several tasks are yet to be allocated, it is not trivial to analyze how many mutually compatible schedulable task sets are formed by the remaining tasks.

\smallskip
\noindent
\revise{\textbf{\emph{Sorting Criterion 2: COMP}.}}
Having established that mutual compatibility between remaining tasks also plays an important role in deciding their schedulability, we propose to use the same sorting criterion as in the first-fit heuristic that works well for non-preemptive tasks~\cite{Baruah:2006}.
That is, we sort the tasks in $\mathcal{T}^{\>*}$  according to a non-decreasing order of their \emph{periods}.
Since tasks with shorter periods will be allocated first, in later iterations of Algorithm~\ref{alg:outer} (the outer layer), when we deal with tasks with longer periods and (likely longer) execution times, there is a high probability that they will \emph{not} be mutually incompatible. 
When our multi-layer optimization uses the above sorting criterion in the middle layer, we term it as \emph{COMP}.

\smallskip
\noindent
\revise{\textbf{Trade-off between cache sensitivity potential and mutual compatibility:}}
Consider that we have allocated two tasks $\tau_1$ and $\tau_2$ on a core. $\tau_1$ has $p_1=10$, $e_1=5$, and $\gamma_1 = 0$ and $\tau_2$ has $p_2=25$, $e_2=5$, and $\gamma_2 = 0$. The core utilization is \SI{70}{\percent}. On the same core, if we want to allocate another task $\tau_3$ with $p_3 = 10$ and $e_3 = 2$, we cannot do it. Now, consider another situation where, instead of $\tau_2$, we have $\tau_4$ that has $p_4 = 10$, $e_4 = 3$, and $\gamma_4 = 0.1$. Now, the core utilization is \SI{80}{\percent}. However, here, we can still add $\tau_3$ to the core without violating schedulability. 
In the first case, the worst-case response time $R_1$ of $\tau_1$ is $10$, equal to its deadline.
Thus, the blocking time for $\tau_1$, \ie, $e_2$, artificially increases the utilization of the core from \SI{70}{\percent} to \SI{100}{\percent}, which is 1.5 times the utilization of $\tau_2$ (\ie, \SI{20}{\percent}). In Figure~\ref{fig:CaseStudy}, we note that the execution time increases by up to \SI{100}{\percent} for the benchmarks we have studied. In the second case, by allocating $\tau_4$, we are compromising \SI{10}{\percent} ($\gamma_4 = 0.1$) of the processor utilization assuming that we can reduce the execution time of $\tau_4$ if we allocate it to another core with more cache partitions.
Note that $\tau_4$ is compatible with $\tau_1$ and $\tau_3$ as they have the same period. 
Again, we see a \emph{trade-off} between considering compatibility and cache sensitivity potential during task allocation. 
We will experimentally evaluate the relative dominance of these two factors in Section~\ref{sec:results}.

\subsection{Algorithm~\ref{alg:outer}: Outer optimization layer}\label{sec:outer} 
In this layer, we focus mainly on cache partitioning, for which we propose an algorithm loosely based on breadth-first search. Each node of the search tree represents a partial solution $\omega$.
A node $\omega$ at a depth $x$ of the search tree comprises the following attributes: (i)~\emph{TaskAlloc} gives the sets of tasks $\{T_1,T_2,\cdots,T_x\}$ allocated to $x$ cores. (ii)~\emph{CachePart} gives the number of cache partitions reserved for tasks allocated to each of the $x$ cores, \ie, $\{\mu_1,\mu_2,\cdots,\mu_x\}$. (iii)~\emph{TasksLeft} is the set of tasks yet to be allocated. (iv)~\emph{CacheLeft} represents the remaining number of cache partitions. (v)~\emph{RemSchedDemand} is the remaining scheduling demand that can be calculated based on \emph{TasksLeft} using Eq.~(\ref{eq:scheddemand}). Clearly, the root node $\omega_{init} \in \Omega_0$ (in line~1) has empty sets in \emph{TaskAlloc} and \emph{CachePart} respectively while \emph{TasksLeft} comprises the complete task set $\mathcal{T}$, \emph{CacheLeft} is equal to $n_p$, and \emph{RemSchedDemand} can be computed as $\sum_{\tau_i \in \mathcal{T}} \widehat{u}_i$. At any time, $\Omega_x$ -- $x$ is the search depth -- will store either the leaf nodes representing a full solution or the parent nodes for which we will explore the child nodes.

\input{Algorithms/algo_2}

Considering that we have $n_c$ cores, the maximum depth of the search tree is $n_c$. In the $x$-th iteration of the for loop in lines~2~-~14, we explore the nodes at depth $x$. $\Omega^*$ (line~3) will store (i)~the leaf nodes only if they represent a full solution and (ii)~non-leaf nodes at depth $x$. 
Using the for loop in lines~4~-~13, we iterate through each node in $\Omega_{x-1}$.
If a node in $\Omega_{x-1}$ already represents a full solution, then it cannot have any valid child nodes and, hence, it is a leaf node. We can add such nodes directly to $\Omega^*$ (lines~12~-~13). Otherwise, we explore the child nodes of a node in $\Omega_{x-1}$.
Note that Algorithm~\ref{alg:outer} does not terminate when it finds a full solution. This is because our goal is also to find the solution that will reserve the minimum number of cache partitions to ensure the schedulability of the task set. Minimizing the number of cache partitions used for real-time tasks maximizes the cache available for soft real-time and best-effort tasks, thus potentially improving the system's overall performance.

For each parent node $\omega$, we can have up to $\omega.$\emph{CacheLeft} number of child nodes. That is, in each iteration of the for loop in lines~6~-~11, we explore a child node (if valid). In the $\mu$-th iteration, we consider that the tasks on the $x$-th core can use $\mu$ cache partitions. We invoke Algorithm~\ref{alg:inner} (the inner layers) in line~7 to obtain a set of tasks $T_x$ to be allocated to the $x$-th core. If $T_x$ is not empty, then we can create a new partial solution $\omega^+$ extending one of the parent nodes (lines~8~-~9). Note that for the parent node, we have the task allocation and cache partitioning until $x-1$ cores and we can now add $T_x$ and $\mu$ to obtain $\omega^+$. We further do some proactive pruning if possible (lines~10~-~11). That is, if $\omega^+$ still have tasks waiting to be allocated while there are no cores or cache partitions left, then we only have a leaf node with an incomplete solution $\omega^+$. There is no point in adding such a node to $\Omega^*$.

After we explore the nodes at depth $x$, we \emph{prune} the search tree further based on a heuristic (in line~14).
We delete a node $\omega' \in \Omega^*$ if it is dominated by another node $\omega \in \Omega^*$.
Here, $\omega$ dominates $\omega'$ if one of the following conditions is satisfied.
\vspace{1mm}
\begin{enumerate}
    \item $\omega.\textit{CacheLeft} > \omega'.\textit{CacheLeft}$ and $\omega.\textit{RemSchedDemand}$ $\leq\>\omega'.\textit{RemSchedDemand}$.
    \vspace{2mm}
    \item $\omega.\textit{CacheLeft} = \omega'.\textit{CacheLeft}$ and $\omega.\textit{RemSchedDemand}$ $< \omega'.\textit{RemSchedDemand}$.
\end{enumerate}
\vspace{1mm}
If $\omega.\textit{CacheLeft} = \omega'.\textit{CacheLeft}$ and $\omega.\textit{RemSchedDemand} =$ $\omega'.\textit{RemSchedDemand}$, we keep just one of them for further exploration and remove the other(s).
The intuition behind the heuristic is as follows: If $\omega'$ already uses an equal or more number of cache partitions than $\omega$ and still have more scheduling demand to meet, then there is a lower probability that a child of $\omega'$ will be a better solution than one of the children of $\omega$.
This pruning is necessary to keep the search tractable. Otherwise, the tree might grow exponentially.
With this pruning heuristic, before exploring the child nodes at depth $x > 1$, we can have only up to $n_p + 2 - x$ parent nodes where each uses a different number of cache partitions.

In the end, in lines~15~-~16, Algorithm~\ref{alg:outer} returns the non-dominated complete solution if it has found one (\ie, if $\Omega_{n_c}$ is not empty)  otherwise, in lines~17~-~18, it returns empty sets for task allocation and cache partitioning, respectively.

\subsection{Complexity analysis} 
In Algorithm~\ref{alg:outer}, we have three nested for-loops. The outer loop (lines~2~-~14) iterates for $n_c$ times. The middle loop (lines~4~-~13) iterates for at most $n_p$ times. This is because, as mentioned earlier, $\Omega_x$ can have a maximum of $n_p + 2 - x$ nodes at the beginning of the $x$-th iteration of the outer loop. The inner loop (lines~6~-~11) also iterates up to $n_p$ times. Hence, the number of times we invoke Algorithm~\ref{alg:inner} from Algorithm~\ref{alg:outer} is upper-bounded by $n_c \cdot n_p^2$. In Algorithm~\ref{alg:inner}, we have only one loop (lines~3~-~6) that iterates at most $n_\tau$ times. Thus, the number of times we invoke the schedulability analysis in the inner layer (line~4) is upper-bounded by \mbox{$n_c \cdot n_p^2\cdot n_\tau$.} 

Further, we evaluate the time complexity of the response time analysis for NP-FP tasks.
The response time analysis of task~$\tau_i$ must cover its entire busy period.  
The busy period $t_i$ of the task~$\tau_i$ must satisfy Equation~(\ref{eq:busyperiod}) and we can write:
\begin{equation} \label{eq:busyperiod_relation}
\begin{aligned}
    t_i &= B_{i,j} + \sum_{\tau_{i'} \in HEP_{i,j}} \left\lceil \frac{t_i}{p_{i'}} \right\rceil e_{i'}
\end{aligned}
\end{equation}
Using the following relation:
\begin{equation*}
 \dfrac{t_i + p_i'}{p_i'}  >
    \left \lceil \dfrac{t_i}{p_i'} \right \rceil 
\end{equation*}
we can upper bound the left-hand side of  Equation~(\ref{eq:busyperiod_relation}) by:
\begin{align}
    t_i < & \; B_{i,j} + \sum_{\tau_{i'} \in HEP_{i,j}}  \frac{t_i+p_{i'}}{p_{i'}}  e_{i'} \nonumber \\
    = & \; B_{i,j} + t_i \sum_{\tau_{i'} \in HEP_{i,j}}  \frac{e_{i'}}{p_{i'}} + 
     \sum_{\tau_{i'} \in HEP_{i,j}}  e_{i'} \nonumber \\
     \leq & \;  \sum_{\tau_{i'} \in T_j} e_{i'} + t_i U_j \nonumber 
\end{align}
where $U_j=\sum_{\tau_{i'} \in T_j} e_{i'}/p_{i'}$ is the total utilization of task set~$T_j$. 
By rearranging the terms, the busy period $t_i$ of each task~$\tau_i \in T_j$ can be upper-bounded as follows:
\begin{equation}
 \forall \tau_i \in T_j: \;   t_i \; < \; \dfrac{\sum_{\tau_{i'} \in T_j} e_{i'}}{1-U_j} \; \leq \; 
 \dfrac{n_\tau \max_{\tau_{i}'\in T_j} e_{i'}}{1-U_j}
\end{equation}
In each iteration of Equation~(\ref{eq:startexec}), we need at most $n_\tau$ operations where the value of $w_i(q)$ increases by at least $\min_{i\in T_j} e_i$ time units (otherwise remains constant) within the busy period.   
The number of operations to compute the task's response time is, hence, upper bounded~by:
\begin{equation}
 \dfrac{\max_{i\in T_j} e_i}{\min_{i\in T_j} e_i} \cdot \dfrac{n_\tau^2}{1-U_j}
\end{equation}

Taking into account the time-complexity of the response time analysis, the overall algorithm's (Algorithm~\ref{alg:inner} and \ref{alg:outer}) asymptotic complexity can be expressed as follows:
\begin{equation}
\mathcal{O} \left (   \dfrac{n_c \cdot n_p^2 \cdot n_\tau^3 }{1-U_j}  \cdot \dfrac{\max_{\tau_i \in T_j} e_i}{\min_{\tau_i \in T_j} e_i} \right).
\end{equation}

Considering that we can put a limit on the utilization of a processing core, \eg, $U_j = \sum_{\tau_i \in T_j} \frac{e_i}{p_i} < 0.99$,
our optimization technique has a pseudo-polynomial time complexity.

%% file: Algorithms/algo_1.tex
\setlength{\textfloatsep}{0pt}
{\SetAlCapFnt{\small}
\SetAlCapNameFnt{\small}
\SetAlFnt{\small}
\begin{algorithm} [t]
    \caption{\revise{\textbf{allocTask}($\ldots$)---Middle and inner layers}}
    \label{alg:inner}
    
    \KwIn{$\big\{\big(p_i,e_i,\epsilon_{i,n_p}\big)|\tau_i \in \mathcal{T}^{\>*}\big\}$\;}
    \KwOut{$T$\;}
    
    $T \gets \varnothing$\;
    $\overline{\mathcal{T}^{\>*}} \gets \textbf{Sort}(\mathcal{T}^*)$ \;
    \tcc{{\color{blue}different sorting criteria can be applied}}%\;
    \For{$\tau_i \in \overline{\mathcal{T}^{\>*}}$}
    {
        \If(\tcp*[f]{inner layer}){$\textbf{\upshape isSchedulable}(T, \tau_i)$}{
            $T.\textbf{append}(\tau_i)$\;
        }
    }
    \KwRet{$T$}\;
\end{algorithm}}

%% file: Algorithms/algo_2.tex
{
\SetAlCapFnt{\small}
\SetAlCapNameFnt{\small}
\SetAlFnt{\small}
\begin{algorithm} [t]
    \caption{Outer optimization layer}
    \label{alg:outer}
    
    \KwIn{$\mathcal{T}$, $n_c$, $n_p$\;}
    \KwOut{$\{T_1,T_2,\cdots, T_{n_c}\}$, $\{\mu_1,\mu_2,\cdots,\mu_{n_c}\}$\;}
    
    $\Omega_0 \gets \{\omega_{init}\}$\;
	\For{$x \gets 1$ \KwTo $n_c$}
        {
	    $\Omega^* \gets \varnothing$\;
	    \For{$\omega \in \Omega_{x-1}$}
            {
	        \eIf{$\omega.TasksLeft \neq \varnothing$}
                {
                    \For{$\mu \gets 1$ \KwTo $\omega.CacheLeft$}
                    {
                        \tcc{{\color{blue}invoke Algorithm~\ref{alg:inner} to allocate tasks to the $x$-th core}}
                        $T_x \gets \textbf{allocTask}(\omega.TasksLeft,\mu)$\;
                        \If{$T_x \neq \varnothing$}
                        {
                            % \tcc{{\color{blue}create a new partial solution by extending the one from the parent node}}
                            \tcc{{\color{blue}extend parent node $\omega$ to a new partial solution $\omega^+$}}
                            $\omega^+ = \textbf{newPartSol}(\omega,T_x,\mu)$\;
                            \tcc{{\color{blue}check if $\omega^+$ can be extended to a full solution}}
                            \If{\textbf{\upshape isProspectiveSolution}$(\omega^+)$}
                            {
                                $\Omega^*.$\textbf{append}($\omega^+$)\;
                            }
                        }
                    }
	        }
    	    {
    	        $\Omega^*.$\textbf{append}($\omega$)\;
    	    }
	    }
	    \tcc{{\color{blue}remove the dominated partial solutions}}
	    $\Omega_x \gets \textbf{removeDominatedPartialSolutions}(\Omega^*)$
	}
	\tcc{{\color{blue}return non-dominated full solution if any}}
	
	\eIf{$\Omega_{n_c} \neq \varnothing$}
        {
	    \KwRet{$\{\Omega_{n_c}[1].TaskAlloc, \Omega_{n_c}[1].CachePart\}$}\;
	}
	{
	    \KwRet{$\{\varnothing, \varnothing\}$}\;    
	}
\end{algorithm}
}

%% file: sections/algex.tex
\section{Illustrative Examples: COMP vs CASE}\label{sec:algex}
Neither heuristics, \emph{COMP} or \emph{CASE}, completely dominate the other in finding schedulable solutions. 
We illustrate this using two examples.
\begin{table}[t]
\scriptsize
\parbox{.495\linewidth}{
\centering
\caption{\emph{CASE} $\prec$ \emph{COMP}}
\label{tab:ex2}
\begin{tabular}{|p{3.5mm}||p{3.5mm}|p{3.5mm}|p{3.5mm}|p{3.5mm}|}
\hline
$\tau_i$ & $\tau_1$ & $\tau_2$ & $\tau_3$ & $\tau_4$ \\ \hline \hline
$p_i$ & 100 & 100 & 150 & 150 \\ \hline \hline
$\epsilon_{i,1}$ & 36 & 75 & 77 & 85 \\ \hline
$\epsilon_{i,2}$ & 35 & 55 & 48 & 82 \\ \hline
$\epsilon_{i,3}$ & 34 & 45 & 35 & 81 \\ \hline  
$\epsilon_{i,4}$ & 34 & 27 & 25 & 79 \\ \hline 
\end{tabular}
}
\hfill
\parbox{.495\linewidth}{
\centering
\caption{\emph{CASE} $\succ$ \emph{COMP}}
\label{tab:ex3}
\begin{tabular}{|p{3.5mm}||p{3.5mm}|p{3.5mm}|p{3.5mm}|p{3.5mm}|}
\hline
$\tau_i$ & $\tau_1$ & $\tau_2$ & $\tau_3$ & $\tau_4$ \\ \hline \hline
$p_i$ & 200 & 200 & 250 & 250 \\ \hline \hline
$\epsilon_{i,1}$ & 35 & 177 & 324 & 65 \\ \hline
$\epsilon_{i,2}$ & 33 & 172 & 178 & 63 \\ \hline
$\epsilon_{i,3}$ & 31 & 168 & 119 & 62 \\ \hline  
$\epsilon_{i,4}$ & 26 & 165 & 80 & 60\\ \hline 
\end{tabular}
}
\vspace{3mm}
\end{table}

\subsection{\emph{COMP} works, \emph{CASE} fails}
Consider an example with $4$ tasks, $4$ cache partitions, and $2$ cores.
Table~\ref{tab:ex2} gives the period of each task and shows how the execution time varies with the number of available cache partitions.
To this task set, we first apply \emph{COMP} to obtain the following schedulable solution:
$$\mu_1 = 2; \hspace{2mm} T_1 = \{\tau_1,\tau_2\};\>\> \mu_2 = 2;\>\> T_2 = \{\tau_3,\tau_4\}.$$
It can be observed that, here, tasks with the same period are co-allocated to a core.

Now, let us try \emph{CASE} on the same example.
At a search depth of $1$, we have two nodes in $\Omega_1$ as follows: 
(i)~$\omega_1$: $\tau_1$ is allocated to the first core and it uses $1$ partition. (ii)~$\omega_2$: $\tau_1$ and $\tau_3$ are allocated to the first core and they share $2$ partitions. 
For $\omega_1$, we have only one task in a core with $(p_1,e_1) = (100,36)$, which is schedulable.
For $\omega_2$, we have two tasks with $(p_1,e_1) = (100,35)$ and $(p_3,e_3) = (150,48)$ and, accordingly, $R_1^{wc}=R_2^{wc} = 83 < 100 < 150$, \ie, the tasks are schedulable. 
Note that \emph{CASE}, in this example, could not co-allocate $\tau_1$ and $\tau_4$ to a core despite having similar cache sensitivity potentials because they are not compatible.
Further, instead of $\tau_2$, $\tau_3$ was selected
after $\tau_1$ in $\omega_2$ because $\gamma_3 = 0.1533$ which is less than $\gamma_2 = 0.28$. 
Following $\omega_1$, we cannot schedule the other $3$ tasks on a single core sharing $3$ remaining partitions because their utilization is greater than 1. Also, with respect to $\omega_2$, $\tau_2$ and $\tau_4$ cannot be co-allocated to a core sharing $2$ partitions because their combined utilization ($\frac{55}{100} + \frac{82}{150}$) is greater than $1$. Hence, \emph{CASE} cannot establish schedulability, unlike \emph{COMP}. 
This shows that there can be a task set for which \emph{CASE} cannot obtain a schedulable solution while \emph{COMP} can.    

\subsection{\emph{CASE} works, \emph{COMP} fails}
Consider another example with task specification in Table~\ref{tab:ex3}.
Let us first study a case where we partition the cache equally, \ie, we provision $2$ cache partitions for the tasks running on each core.
Corresponding to this, the execution time of each task gets fixed.
Thus, we have to schedule 4 tasks on 2 cores with the following specification:
$$(p_1,e_1) = (200,33); \hspace{3mm} (p_2,e_2) = (200,172);$$
$$(p_3,e_3) = (250,178); \hspace{3mm} (p_4,e_4) = (250,63).$$
We can write the utilization of the tasks as follows:
$$u_1 = 0.165; \hspace{2mm}u_2 = 0.86; \hspace{2mm}u_3 = 0.712; \hspace{2mm}u_4 = 0.252.$$
It is obvious that no task can be allocated together with $\tau_2$ to a core because it will lead to a core utilization greater than 1.
Alternatively, if we allocate the other $3$ tasks ($\tau_1$, $\tau_3$, and $\tau_4$) to a core, the total utilization will become $0.165 + 0.712 + 0.252 = 1.129 > 1$, which is again infeasible. Hence, this task set is not schedulable if we consider equal cache partitioning.

Now, when we apply \emph{CASE} to this example, we get the following schedulable solution: 
$$\mu_1 = 3; \hspace{2mm} T_1 = \{\tau_1,\tau_3,\tau_4\};\>\> \mu_2 = 1;\>\> T_2 = \{\tau_2\}.$$
This example, therefore, illustrates that exploring the design space for cache partitioning together with task allocation and scheduling (using our proposed framework) can improve the likelihood of establishing the schedulability of a task set, which is the main motivation behind this work. 

Further, we apply \emph{COMP} to the same example.
We have two nodes in $\Omega_1$ at a search depth of $1$.
(i)~$\omega_1$: $\tau_1$ and $\tau_4$ are allocated to the first core and share $1$ partition. Thus, we get $R_1^{wc} = R_4^{wc} = 100 < 200 < 250$.
(ii)~$\omega_2$: $\tau_1$ and $\tau_2$ are allocated to the first core and share $3$ partitions. Here, we get $R_1^{wc} = R_2^{wc} = 199 < 200$.
In case of $\omega_1$, the two remaining tasks are $\tau_2$ and $\tau_3$. They are not co-schedulable on a core even with 3 remaining partitions because their combined utilization is $\frac{168}{200} + \frac{119}{250}>1$.
In case of $\omega_2$, the remaining tasks are $\tau_3$ and $\tau_4$. They cannot be mapped to one core and share $1$ partition because their utilizations add up to $\frac{324}{250} + \frac{65}{250}>1$.
Hence, with \emph{COMP}, we could not obtain a solution.
Thus, this example shows that \emph{CASE} can obtain a schedulable solution in some cases where \emph{COMP} cannot.  

%% file: sections/results.tex
\section{Experimental Results}\label{sec:results}
In this section, we will first explain different scenarios for the experiments, and then, we will present the experimental results and discuss our observations.
\subsection{Design of experiments}
\noindent\textbf{Benchmark study:}
We first experimentally investigate the relation between a task’s execution time and available cache size for real-world benchmarks.
These benchmarks include computer vision applications from  Georgia Tech Smoothing and Mapping (\eg,~structure from motion)~\cite{gtsam}, VLFeat (\eg,~k-means clustering)~\cite{vlfeat}, and OpenCV (\eg,~letter recognition) \cite{opencv_library} as well as motion planning algorithms from Open Motion Planning Library (\eg,~geometric car planning)~\cite{benchmarking-motion-planning-algorithms}.\footnote{The complete set of benchmark task profiles are available in the appendix.}
We simulate the execution of each benchmark using \revise{the} \emph{Cachegrind}\footnote{\url{https://valgrind.org/info/tools.html/\#cachegrind}} instrumentation tool and measure the number of instructions executed $I$, data cache hits $DH(\mu)$ and data cache~misses $DM(\mu)$ for data cache sizes $\mu$ between $1$ and~$8192$~KB. 
In Cachegrind, we have specified an 8-way set-associative last-level cache and a cache line size of 64~bytes. Note that Cachegrind only supports the \emph{Least Recently Used} replacement~policy, and the number of sets is restricted to be a power of two.

We estimate the execution time $\mathcal{E}(\mu)$ of a given benchmark program for an available cache of size $\mu$ as follows:
\begin{equation}
\label{eq:exec_time}
\mathcal{E}(\mu) = I \cdot CPI + DM(\mu) \cdot  MP  + DH(\mu) \cdot  HP,
\end{equation}
where $CPI$ is the number of clock cycles per instruction, $MP$ and $HP$ are the cache miss and hit penalties, respectively, in terms of the number of processor cycles (all parameters are platform-specific). 
We assume a superscalar processor that can execute $1/CPI {=} 2$ instructions per cycle and last-level cache hit and miss penalties of $HP{=}20$ and $MP{=}200$~cycles.

The proposed technique requires as input the execution time function for different cache sizes. We note that any method can be used to obtain it, including WCET analysis \revise{tools}~\cite{Ballabriga:2010,Ferdinand:2004,Chronos} or measurement-based approaches~\cite{Bernat:2002}.
Our Cachegrind-based technique enables a fine-granular characterization of the dependence of execution time on the number of available cache partitions. The slowdown profiles are compatible with those obtained by both simulation methods and real observations in previous works~\cite {MDBCCP:13,XPCLLLL:19,Kwon:2021,Lesage:2015,KWCFAS:17}.
\revise{In our experiments, it took 1 to 5 hours to obtain the execution time of a task for a particular cache-partition assignment. We note that such profiles need to be obtained only once, and the WCET profile for each task can be obtained independently and in parallel with other tasks.}

We experiment with a wide variety of benchmarks (in terms of how much slowdown they can suffer) and show how varying slowdown affects the results.
Although more than $50$ benchmark programs have been analyzed, for our further experiments, we use the four most representative ones, \ie, with a good spread of maximum slowdown values. 
The \emph{slowdown} in the execution of these four benchmarks as a function of available cache size between 64~KB and 2048~KB is shown in Figure~\ref{fig:CaseStudy}.

\smallskip
\noindent\textbf{Synthetic slowdown profiles:} Besides the slowdown profiles of benchmark programs, we use $8$ synthetic profiles where the execution time decreases exponentially with the number of cache partitions. We can write these profiles as follows:
\begin{equation*}
\begin{aligned}
    &\hspace{2cm}\frac{\epsilon_{i,\mu}}{\epsilon_{i,n_p}} = \frac{\text{exp}(-\mu\alpha)}{\text{exp}(-n_p\alpha)}\>\>\>\> \text{where,}\\
    &\alpha \in \{0,0.023,0.036,0.045,0.052,0.058,0.067,0.0743\}.
\end{aligned}
\end{equation*}
Recall that $\epsilon_{i,\mu}$ is the execution time of the task when it can use only $\mu$ cache partitions and $\epsilon_{i,n_p}$ is the execution time when it can use all cache partitions.
Note that the values of $\alpha$ are selected to obtain maximum slowdowns of 1, 2, 3, 4, 5, 6, 8, and 10 when $n_p = 32$.
For $n_p = 16$, the maximum slowdown values are 1, 1.4, 1.7, 2, 2.2, 2.4, 2.7, and 3. 
We note that such slowdowns can occur in practice for real workloads as reported in~\cite{farshchi_et_al:LIPIcs:2018:9001, Roozkhosh:2020}.
We denote the profiles as $\mathcal{P}^{syn}_1,\mathcal{P}^{syn}_2, \cdots, \mathcal{P}^{syn}_8$ with respect to the values of $\alpha$ in ascending order.

\smallskip
\noindent
\textbf{Cache configurations:} Inspired by MPSoC architectures with different cache sizes, we consider two scenarios
\mbox{$\mathbb{AR}$-I} and \mbox{$\mathbb{AR}$-II}, with four cores and a maximum number of partitions $n_p = 16$ and $n_p = 32$, respectively.

\smallskip
\noindent
\textbf{Test case generation:} For generating a test case,
{we consider a system with 4 cores and 40 tasks.}
We use~\cite{Emberson:2010} to randomly synthesize base utilization of tasks for a target utilization $U_{tar}$ where $U_{tar} = \sum_{\tau_i \in \mathcal{T}} \widehat{u}_i$. For each $U_{tar} \in \{1.0,1.1, \cdots,3.9,4\}$, we generate 100 task sets. To each task in a task set, we assign a slowdown profile (variation of execution time/utilization with available cache) either similar to an evaluated benchmark program---linearly interpolating for the non-available cache sizes---or generated synthetically.

\smallskip
\noindent
\textbf{Selections of slowdown profiles:}
We take three different sets of slowdown profiles.
(i)~$\mathbb{SD}$-B comprises slowdown profiles of four benchmark programs as depicted in Figure~\ref{fig:CaseStudy}.
(ii)~$\mathbb{SD}$-S1 comprises synthetic slowdown profiles $\mathcal{P}^{syn}_1$, $\mathcal{P}^{syn}_2$, $\mathcal{P}^{syn}_3$, $\mathcal{P}^{syn}_4$, $\mathcal{P}^{syn}_5$, and $\mathcal{P}^{syn}_6$.
(iii)~$\mathbb{SD}$-S2 also includes synthetic profiles with high slowdown, specifically:
$\mathcal{P}^{syn}_1$, $\mathcal{P}^{syn}_2$, $\mathcal{P}^{syn}_4$, $\mathcal{P}^{syn}_6$, $\mathcal{P}^{syn}_7$, and $\mathcal{P}^{syn}_8$. 

\smallskip
\noindent
\textbf{Selections of task periods:}
We consider two sets of values from which we choose task periods: (i)~$\mathbb{WD}$: This set comprises a wider range of values. Each task $\tau_i \in \mathcal{T}$ is randomly assigned a period from $\{5, 10, 20, 40, 60, 80,$ $100\}$. Further, we do not limit the task utilization. Thus, in this case, two tasks with a wide gap in their periods, \eg, 5~ms and 100~ms, can easily be incompatible for co-scheduling. (ii)~$\mathbb{SH}$: This set comprises a shorter range of values, \ie, we select task periods from $\{10, 15, 20, 25\}$. Moreover, we limit the base utilization of a task to $0.2$ in this case. 
Considering that the gap between two task periods is not very wide and the utilization of a low-priority task may not be very high when we select periods from $\mathbb{SH}$, the likelihood of two tasks not being schedulable together in a core is low.

\smallskip
\noindent
\textbf{Scenarios for experiments:}
In summary, 
we consider two cache configurations, 
%We have two MPSoC architectures,
$\mathbb{AR}$-I and $\mathbb{AR}$-II, with 16 and 32 partitions, respectively. %differing in the cache size, \ie, $n_p = 16$ and $n_p = 32$, respectively.
Further, we have three selections of slowdown profiles, \textit{i.e.}, $\mathbb{SD}$-B, $\mathbb{SD}$-S1, and $\mathbb{SD}$-S2.
Also, we have two sets of task periods, \textit{i.e.}, $\mathbb{WD}$ and $\mathbb{SH}$.
In combination, we have $12$ different scenarios as given in Tables~\ref{tab:results_npfp}-\ref{tab:results_pedf}.

\smallskip
\noindent
\textbf{Baseline algorithms:} 
To the best of our knowledge, no previous algorithm has been specifically proposed for co-optimizing cache partitioning and task allocation under the NP-FP scheduling policy. In our experiments, we compare \emph{CASE} and \emph{COMP} against three state-of-the-art algorithms, \ie, \emph{CaM}~\cite{XPCLLLL:19}, \emph{IA}$^3$~\cite{Paolieri:2011}, and \emph{PDPA}~\cite{Berna:2012}.
Since these algorithms were initially proposed for various scheduling policies (\emph{CaM} for P-EDF, \emph{IA}$^3$ and \emph{PDPA} for NP-EDF), we have adapted them preserving their allocation strategy, but using the NP-FP schedulability test.
Conversely, in Section~\ref{sec:results_edf}, \emph{CASE} and \emph{COMP} have been modified to suit P-EDF and NP-EDF.
In the following, we first present the evaluation under NP-FP, which is the primary goal of the paper.
Then, we demonstrate the flexibility of the optimization framework by comparing it with the scheduling policies originally targeted by each baseline algorithm.
Except for considering different schedulability tests, the baseline algorithms were not otherwise modified.\footnote{Similarly to \eg,~\cite{XPCLLLL:19}, when considering preemptive scheduling, we assume that CRPD is already included in the WCET of tasks.}

\subsection{Experimental results under NP-FP}
\label{sec:results_npfp}

The overall schedulability results under the NP-FP scheduling policy are reported in Table~\ref{tab:results_npfp}.
Besides, we illustrate the results of six representative scenarios in Figure~\ref{fig:sched_results_selected}.\footnote{All schedulability plots are available in the appendix.}

\input{tables/results_npfp}

\input{figures/sched_results_selected}

\smallskip
\noindent
\textbf{Comparison among baselines:}
\revise{
Both \emph{IA}$^3$ and \emph{PDPA} are developed for NP-EDF, where blocking from a lower-priority task influences the schedulability, which is also the case for NP-FP. 
\emph{PDPA} tends to allocate tasks with similar periods to the same core, while \emph{IA}$^3$ considers cache sensitivity as a major deciding factor.
Although we expected \emph{PDPA} to perform well for task sets with a smaller variation in the cache sensitivity, our experiments suggest that \emph{IA}$^3$ performs significantly better (at least 24\% more schedulable task sets) than \emph{PDPA} in each scenario, as shown in Table~\ref{tab:results_npfp}. 

We have identified the following shortcomings of \emph{PDPA} that might affect its performance when applied to the task sets we generate.
First, \emph{PDPA} attempts to initially assign one critical task to each core by selecting among the tasks with high utilization and low cache variability. After the critical tasks are assigned, other tasks can only be mapped to the core hosting a critical task with a higher or equal period. However, the algorithm does not guarantee that the task with the largest period is always selected as a critical task. In such cases, the tasks with periods higher than all critical tasks cannot be scheduled to any core.
Our implementation of \emph{PDPA} tries to avoid this situation by first selecting the task with the highest period as a critical task and then assigning it to the first core before selecting other critical tasks. 
Second, \emph{PDPA} has a hard constraint 
(\ie, line 9 of Algorithm 2 in \cite{Berna:2012})
to ensure that the periods of critical tasks are as far as possible from each other. \cite{Berna:2012} reports that $\delta = \frac{P_{max} - P_{min}}{M} \times \frac{\Delta=90}{100}$ produces the best performance in the experiments presented in that paper. %Unfortunately,
The parameter $\Delta = 90$ does not work with our task sets since, in some cases, no tasks can satisfy the constraint, and thus, the number of critical tasks is smaller than the number of cores. In our experiments, we therefore reduced $\Delta = 50$ so that one critical task is assigned to each core. 
Third, after generating an initial task allocation based on the critical tasks,  
\emph{PDPA} only remaps selected tasks if a core is not schedulable even with the full cache.
However, the algorithm does not check whether the total cache allocation is larger than the available cache size. 
Additionally, if every core is schedulable with the full cache, \emph{PDPA} will not try to reduce the cache allocation to obtain a valid solution.
As reported in~\cite{Berna:2012}, when task sets are restrictively generated to achieve minimum execution time for each task with 1 to 4 cache partitions, \emph{PDPA} might perform well.
However, we do not impose such restrictions while generating our task sets.

Different from \emph{IA}$^3$ and \emph{PDPA}, \emph{CaM} is developed for P-EDF, and it allocates tasks with similar slowdown profiles to the same core using \emph{k-means} clustering.
It does not consider non-preemptive blocking and might allocate mutually incompatible tasks to the same core.
As a result, \emph{CaM} does not perform well when task periods are generated from $\mathbb{WD}$, as shown in Table~\ref{tab:results_npfp}.
However, when the task periods are chosen from $\mathbb{SH}$, \emph{CaM} can schedule 2.9\% more task sets than \emph{IA}$^3$.
Moreover, \emph{CaM} performs better than other baselines (by 27.5\%) when the cache sensitivity of tasks is high, \eg, $\mathbb{AR}$-II $+$ $\mathbb{SD}$-S1 and $\mathbb{AR}$-II $+$ $\mathbb{SD}$-S2. 
It demonstrates that \emph{CaM} can better explore cache sensitivities of tasks because of its sophisticated clustering technique.
}

\smallskip
\noindent
\textbf{\emph{COMP} vs baselines:} 
In all scenarios, \emph{COMP} performs better than the three baseline algorithms, \emph{IA}$^3$, \emph{PDPA}, and \emph{CaM}, in terms of schedulability (see Table~\ref{tab:results_npfp} and Figure~\ref{fig:sched_results_selected}).
On average, \emph{COMP} performs 13.5\% better than the best-performing baseline.
When the tasks have low cache sensitivities (a maximum slowdown of 3x in the scenarios involving \mbox{$\mathbb{AR}$-I}), \emph{COMP} improves schedulability over the best-performing baseline by 6.1\% on average (Table~\ref{tab:results_npfp}).
We get a minimum gain of 1.6\% when the maximum slowdown is 1.97x (\ie, using $\mathbb{SD}$-B) and task periods are selected from the wider range of values (\ie, $\mathbb{WD}$).     
Conversely, when the tasks have high cache sensitivities (up to 10x slowdown with $\mathbb{SD}$-S2 and $\mathbb{AR}$-II), \emph{COMP} can schedule 97.3\% (for $\mathbb{SH}$) and 233.6\% (for $\mathbb{WD}$) more task sets. 
\revise{
Although \emph{COMP} does not cluster tasks with similar cache sensitivities, it can improve schedulability significantly compared to the baselines.
We believe the main reason is the thorough design space exploration by the proposed multi-layer optimization framework.
In particular, we interleave the allocation of cache partitions and tasks, and hence, we explore adequate combinations of cache partitions and tasks for each core.
At the same time, we keep the search tractable using the proposed pruning strategy explained in Section~\ref{sec:outer}.
}

\smallskip
\noindent
\textbf{\emph{CASE} vs baselines:} 
\revise{
\emph{CASE} considers cache sensitivity for allocating tasks and cache partitions to cores. 
Considering all scenarios in our experiments, it performs 5.5\% better than the best-performing baseline on average.
However, it performs worse than \emph{IA}$^3$ when task periods are selected from $\mathbb{WD}$ (\ie, the ratio between two task periods can be as high as 20) and tasks have low cache sensitivities (\ie, up to 3x slowdown).
For example, \emph{CASE} performs 16.3\% worse than \emph{IA}$^3$ in scenario $\mathbb{AR}$-I $+$ $\mathbb{WD}$ $+$ $\mathbb{SD}$-B, as shown in Table~\ref{tab:results_npfp}.
For such task sets, the gain for clustering tasks with similar cache sensitivities is limited and easily dominated by mutual incompatibility issues.
On contrary, for the scenarios where tasks can have high cache sensitivity (maximum slowdown larger than 3x), \emph{CASE} performs better than \emph{CaM} and \emph{IA}$^3$ (see Table~\ref{tab:results_npfp} and Figures~\ref{fig:ar1_sh_s2}-\ref{fig:ar2_wd_s2}).
For example, it can schedule 140.6\% more task sets than the best-performing baseline (\ie, \emph{CaM}) in scenario $\mathbb{AR}$-II $+$ $\mathbb{SH}$ $+$ $\mathbb{SD}$-S2.
Besides the more in-depth design space exploration achieved by the proposed optimization framework, our proposed metric (\ie, cache sensitivity potential) also enhances the performance of \emph{CASE}.
Using this metric, the algorithm has more potential to explore solutions where tasks' utilizations are close to their base utilization.
}

\smallskip
\noindent
\textbf{\emph{COMP} vs \emph{CASE}:} In our experiments, \emph{CASE} has higher schedulability than \emph{COMP} when both cache sensitivity of tasks is high and task periods are in a shorter range, \ie, $\mathbb{AR}$-II $+$ $\mathbb{SH}$ (see Table~\ref{tab:results_npfp} and Figures~\ref{fig:ar2_sh_b}-\ref{fig:ar2_sh_s1}).
Here, with $\mathbb{SD}$-S2 (up to 10x slowdown), \emph{CASE} schedules 21.94\% more task sets than \emph{COMP}
(Table~\ref{tab:results_npfp}).
In scenarios with low cache sensitivity and shorter range of task periods (\eg, $\mathbb{SH}$ $+$ $\mathbb{SD}$-B), the average difference between \emph{CASE} and \emph{COMP} is smaller.
In scenarios with wider period range ($\mathbb{WD}$), \emph{COMP} performs at least 16\% better than \emph{CASE}
(see Table~\ref{tab:results_npfp} and Figures~\ref{fig:ar1_wd_b}, \ref{fig:ar1_wd_s1}, \ref{fig:ar2_wd_s2}). 
The maximum improvement is 21.38\% in scenario $\mathbb{AR}$-I $+$ $\mathbb{WD}$ $+$ $\mathbb{SD}$-B
(see Table~\ref{tab:results_npfp} and Figure~\ref{fig:ar1_wd_b}).
Overall, in our experiments, \emph{COMP} performs 7.56\% better than \emph{CASE}.
Therefore, an optimal use of our framework should adopt
\emph{CASE} when tasks have high cache sensitivities and their periods are in a shorter range. Otherwise, it should use \emph{COMP}. 
If we run \emph{COMP} or \emph{CASE} based on the characteristics of the task set---as identified from the results---our optimization framework can schedule 15.2\% more task sets than the baselines.
It is also trivial to run both \emph{COMP} and \emph{CASE} in parallel for a task set and select the highest schedulability.

\begin{figure}[t]
    \centering
        \includegraphics[width=0.9\columnwidth]{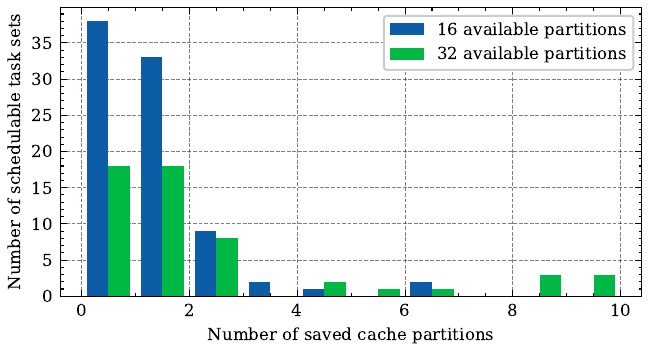}
    \vspace{-2mm}
    \caption{Proposed schemes reserve less cache than baselines.}
    % \vspace{-3mm}
\label{fig:cachesaved}
\end{figure}

\smallskip
\noindent
\textbf{Minimize cache reservation:} Besides improving schedulability, our proposed schemes use less cache compared to baseline algorithms for almost all task sets.
Let us denote $\mu_\mathcal{T}(\mathcal{A}) = \sum_{C_j \in \mathcal{C}} \mu_j$
the total number of cache partitions reserved for a task set $\mathcal{T}$ by an algorithm $\mathcal{A}$.
The minimum between $\mu_\mathcal{T}(\textit{COMP})$ and $\mu_\mathcal{T}(\textit{CASE})$ is denoted by $\mu_\mathcal{T}(\textit{PROP})$. For the baseline algorithms,
after the algorithm terminates and
if the task set is schedulable, we go through the task allocation on each core and try to reduce the number of cache partitions reserved for the tasks without jeopardizing their schedulability. After this post-processing step, 
$\mu_\mathcal{T}(\textit{BASE})$ is 
the minimum cache partitions used among all baseline algorithms.
If a task set $\mathcal{T}$ is deemed schedulable by both a proposed strategy and a baseline algorithm, we compute the number of cache partitions saved as $\mu_{save} = \mu_\mathcal{T}(\textit{BASE}) - \mu_\mathcal{T}(\textit{PROP})$.
On average, with $\mathbb{AR}$-I (16 available partitions), we can save 0.73 partitions, while with $\mathbb{AR}$-II (32 available partitions), we can save 3.68 partitions. Thus, we can save ca. 8.0\% of the available partitions, which is significant.
Non-real-time tasks can benefit from this saving without jeopardizing real-time performance.
Figure~\ref{fig:cachesaved} shows histograms for two scenarios with benchmarks profiles $\mathbb{SD}$-B, wider periods $\mathbb{WD}$, and 16/32 cache partitions.
For certain task sets, we can save up to 5 (10) cache partitions out of 16 (32) available partitions, which is more than 31\% of the cache size.

\subsection{Experimental results under P-EDF and NP-EDF}
\label{sec:results_edf}

In Table~\ref{tab:results_npedf} and \ref{tab:results_pedf}, we evaluate the flexibility of our framework using the
scheduling policies originally targeted by the baseline algorithms (\ie, P-EDF for \emph{CaM}, NP-EDF for \emph{IA$^3$} and \emph{PDPA}).\footnote{Graphs for all results are available in the appendix.}
We modified our framework to use the appropriate schedulability test---NP-EDF test~\cite{jeffay1991non} (used by \cite{Paolieri:2011, Berna:2012}), P-EDF test~\cite{liu1973scheduling} (used by \cite{XPCLLLL:19})---in the inner layer. The middle and outer layers have not been changed.

Results for NP-EDF (Table~\ref{tab:results_npedf}) show a similar trend as NP-FP (Table~\ref{tab:results_npfp}):
\emph{COMP} performs better than \emph{IA$^3$} and \emph{PDPA} for all scenarios, while \emph{CASE} performs better than the baselines when tasks have higher cache sensitivities or shorter range of periods.
On average, \emph{COMP} and \emph{CASE} improve schedulability by 17.1\% and 11.4\%, respectively, over the best-performing baseline (\ie, \emph{IA$^3$}).
For each scenario, if we consider our better-performing algorithm (\emph{CASE} or \emph{COMP}) and compare it against the better-performing baseline, the average improvement is 19.2\%.
The maximum improvement of 261\% is achieved by \emph{CASE} in scenario $\mathbb{AR}$-II $+$ $\mathbb{SH}$ $+$ $\mathbb{SD}$-S2.

\input{tables/results_npedf}

Table~\ref{tab:results_pedf} reports the schedulability results under P-EDF, which was originally targeted by \emph{CaM} \cite{XPCLLLL:19}. The results show a different trend than under non-preemptive scheduling: \emph{CASE} performs better than \emph{CaM} in all scenarios and better than \emph{COMP} in most cases.
Unlike under non-preemptive scheduling, a longer-executing, lower-priority (or longer period) task cannot block a shorter-deadline higher-priority (or shorter period) task. Therefore, task periods and mutual compatibility play a less important role than cache sensitivity. 
Moreover, the overall improvement (8.7\% on average) achieved by the proposed algorithms under P-EDF is smaller than NP-FP and NP-EDF. In fact, in \emph{COMP} and \emph{CASE}, the deciding factors (\ie, cache sensitivity and task period) used in the middle layer are specifically proposed for non-preemptive tasks.
In future work, we would like to investigate whether a new task selection mechanism for P-EDF could produce better results.

\input{tables/results_pedf}

\subsection{Running Time}
\input{tables/rumtime}
We implemented all the algorithms under comparison using Python 3.10 and
conducted our experiments on a Linux server equipped with Intel Xeon Gold 6254 CPU (3.10 GHz). Table~\ref{tab:runtime} presents the average and maximum running time (in seconds) required by the proposed optimization framework and the baseline algorithms.
The running time does not include the time for deriving the execution time function for each task. 
For our experiments under NP-FP, \emph{COMP} and \emph{CASE} required an average running time of less than \SI{1}{\second} and \SI{2}{\second} with 16 and 32 cache partitions, respectively, which are comparable with \emph{IA$^3$} and faster than \emph{CaM}.
For the scenario under NP-EDF, the proposed framework ran slower than \emph{IA$^3$} and \emph{PDPA}, with infrequent spikes due to a combination of multiple iterations and the slower schedulability test.
P-EDF running times are shorter than \mbox{NP-FP} and NP-EDF because of the faster utilization bound test.

Furthermore, we also analyzed the running time scalability of the framework (for NP-FP) using hypothetical test cases with 16 cores, 128 cache partitions, and 160 tasks, which took an average running time of \SI{22}{\minute}s.
Note that our current implementation only utilizes one CPU core for one problem instance.
Exploiting the inherent parallelism in the search in the outer layer can speed up the optimization manifolds.

%% file: tables/results_npfp.tex
\begin{table}[t]
  \begin{center}
  \scriptsize
    \vspace{1mm}
    \caption{\small{Total number of schedulable task sets in each scenario using different algorithms under NP-FP. In \textbf{Bold} (\textit{italic}), the results of the best-performing algorithm (best-performing baseline).}}
    \label{tab:results_npfp}
     \renewcommand{\arraystretch}{1.2}
    \begin{tabular}{|c||c|c|c|c|c|}
    \hline
    \textbf{Scenario} & \emph{COMP} & \emph{CASE} & \emph{IA}$^3$ & \emph{PDPA} & \emph{CaM}  \\ \hline \hline
     $\mathbb{AR}$-I $+$ $\mathbb{SH}$ $+$ $\mathbb{SD}$-B & \textbf{1954} & 1889 & \textit{1887} & 724 & 1768 \\ \hline
     $\mathbb{AR}$-I $+$ $\mathbb{SH}$ $+$ $\mathbb{SD}$-S1 & \textbf{1558} & 1523 & \textit{1450} & 863 & 1408 \\ \hline
     $\mathbb{AR}$-I $+$ $\mathbb{SH}$ $+$ $\mathbb{SD}$-S2 & \textbf{1302} & 1280 & \textit{1153} & 736 & 1142 \\ \hline
     $\mathbb{AR}$-I $+$ $\mathbb{WD}$ $+$ $\mathbb{SD}$-B & \textbf{1981} & 1632 & \textit{1950} & 608 & 169 \\ \hline
     $\mathbb{AR}$-I $+$ $\mathbb{WD}$ $+$ $\mathbb{SD}$-S1 & \textbf{1564} & 1335 & \textit{1475} & 615 & 140 \\ \hline
     $\mathbb{AR}$-I $+$ $\mathbb{WD}$ $+$ $\mathbb{SD}$-S2 & \textbf{1293} & 1101 & \textit{1185} & 480 & 78 \\ \hline
     $\mathbb{AR}$-II $+$ $\mathbb{SH}$ $+$ $\mathbb{SD}$-B & 2356 & \textbf{2434} & 2064 & 1357 & \textit{2254} \\ \hline
     $\mathbb{AR}$-II $+$ $\mathbb{SH}$ $+$ $\mathbb{SD}$-S1 & 760 & \textbf{832} & 501 & 379 & \textit{584} \\ \hline
     $\mathbb{AR}$-II $+$ $\mathbb{SH}$ $+$ $\mathbb{SD}$-S2 & 515 & \textbf{628} & 154 & 124 & \textit{261} \\ \hline
     $\mathbb{AR}$-II $+$ $\mathbb{WD}$ $+$ $\mathbb{SD}$-B & \textbf{2348} & 2014 & \textit{2050} & 1105 & 287 \\ \hline
     $\mathbb{AR}$-II $+$ $\mathbb{WD}$ $+$ $\mathbb{SD}$-S1 & \textbf{801} & 664 & \textit{521} & 253 & 20 \\ \hline
     $\mathbb{AR}$-II $+$ $\mathbb{WD}$ $+$ $\mathbb{SD}$-S2 & \textbf{497} & 407 & \textit{149} & 64 & 1   \\ \hline
    \end{tabular}
  \end{center}
  \vspace{1.5mm}
\end{table}

%% file: figures/sched_results_selected.tex
\begin{figure*}[ht]
    \centering
    \begin{subfigure}{.33\textwidth}
      \centering
      \includegraphics[width=0.95\linewidth]{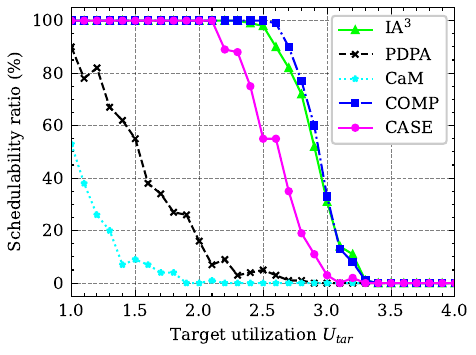}
      \caption{$\mathbb{AR}$-I $+ \mathbb{WD} + \mathbb{SD}$-B.}
      \label{fig:ar1_wd_b}
    \end{subfigure}%
    \begin{subfigure}{.33\textwidth}
      \centering
      \includegraphics[width=0.95\linewidth]{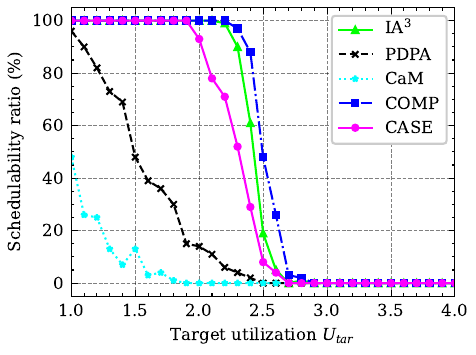}
      \caption{$\mathbb{AR}$-I $+ \mathbb{WD} + \mathbb{SD}$-S1.}
      \label{fig:ar1_wd_s1}
    \end{subfigure}%
    \begin{subfigure}{.33\textwidth}
      \centering
      \includegraphics[width=0.95\linewidth]{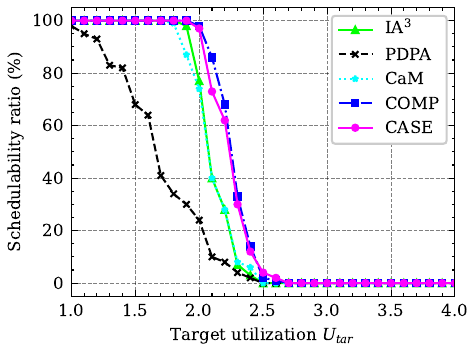}
      \caption{$\mathbb{AR}$-I $+ \mathbb{SH} + \mathbb{SD}$-S2.}
      \label{fig:ar1_sh_s2}
    \end{subfigure}%
    \\
     \begin{subfigure}{.33\textwidth}
      \centering
      \includegraphics[width=0.95\linewidth]{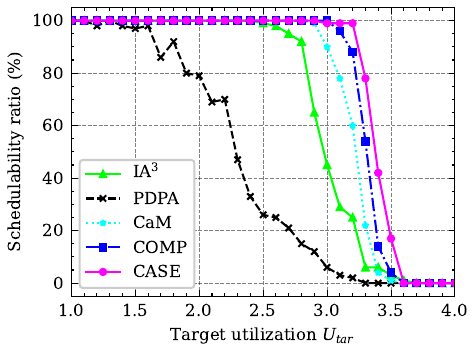}
      \caption{$\mathbb{AR}$-II $+ \mathbb{SH} + \mathbb{SD}$-B.}
      \label{fig:ar2_sh_b}
    \end{subfigure}%
    \begin{subfigure}{.33\textwidth}
      \centering
      \includegraphics[width=0.95\linewidth]{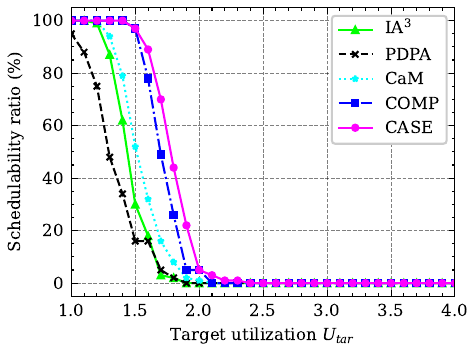}
      \caption{$\mathbb{AR}$-II $+ \mathbb{SH} + \mathbb{SD}$-S1.}
      \label{fig:ar2_sh_s1}
    \end{subfigure}%
    \begin{subfigure}{.33\textwidth}
      \centering
      \includegraphics[width=0.95\linewidth]{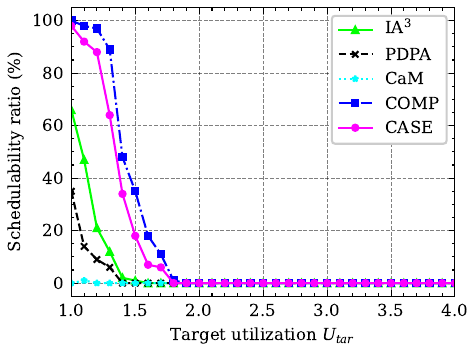}
      \caption{$\mathbb{AR}$-II $+ \mathbb{WD} + \mathbb{SD}$-S2.}
      \label{fig:ar2_wd_s2}
    \end{subfigure}%
    \caption{Schedulability ratio (\textit{i.e.}, $\frac{\text{\#schedulable task sets}}{\text{\#total task sets}} \times 100$ \%) of proposed schemes and baseline algorithms under NP-FP.}
    \label{fig:sched_results_selected}
    \vspace{-3mm}
\end{figure*}

%% file: tables/results_npedf.tex
\begin{table}[t]
  \begin{center}
  \scriptsize
    %\vspace{3mm}
    \caption{\small{Total number of schedulable task sets in each scenario using different algorithms under NP-EDF. In \textbf{Bold} the results of the best-performing algorithm.}}
    \label{tab:results_npedf}
     \renewcommand{\arraystretch}{1.2}
    \begin{tabular}{|c||c|c|c|c|}
    \hline
    \textbf{Scenario} & \emph{COMP} & \emph{CASE} & \emph{IA}$^3$ & \emph{PDPA}  \\ \hline \hline
    $\mathbb{AR}$-I $+$ $\mathbb{SH}$ $+$ $\mathbb{SD}$-B & \textbf{2096} & 2089 & 2075 & 841  \\ \hline
    $\mathbb{AR}$-I $+$ $\mathbb{SH}$ $+$ $\mathbb{SD}$-S1  & \textbf{1695} & 1684 & 1586 & 937  \\ \hline
    $\mathbb{AR}$-I $+$ $\mathbb{SH}$ $+$ $\mathbb{SD}$-S2  & 1447 & \textbf{1455} & 1294 & 804  \\ \hline
    $\mathbb{AR}$-I $+$ $\mathbb{WD}$ $+$ $\mathbb{SD}$-B   & \textbf{2109} & 1803 & 2075 & 767  \\ \hline
    $\mathbb{AR}$-I $+$ $\mathbb{WD}$ $+$ $\mathbb{SD}$-S1  & \textbf{1699} & 1467 & 1572 & 688  \\ \hline
    $\mathbb{AR}$-I $+$ $\mathbb{WD}$ $+$ $\mathbb{SD}$-S2  & \textbf{1413} & 1237 & 1272 & 572  \\ \hline
    $\mathbb{AR}$-II $+$ $\mathbb{SH}$ $+$ $\mathbb{SD}$-B  & 2522 & \textbf{2711} & 2259 & 1514 \\ \hline
    $\mathbb{AR}$-II $+$ $\mathbb{SH}$ $+$ $\mathbb{SD}$-S1 & 925  & \textbf{972}  & 584  & 410  \\ \hline
    $\mathbb{AR}$-II $+$ $\mathbb{SH}$ $+$ $\mathbb{SD}$-S2 & 677  & \textbf{769}  & 213  & 139  \\ \hline
    $\mathbb{AR}$-II $+$ $\mathbb{WD}$ $+$ $\mathbb{SD}$-B  & \textbf{2519} & 2237 & 2172 & 1350 \\ \hline
    $\mathbb{AR}$-II $+$ $\mathbb{WD}$ $+$ $\mathbb{SD}$-S1 & \textbf{914}  & 775  & 590  & 318  \\ \hline
    $\mathbb{AR}$-II $+$ $\mathbb{WD}$ $+$ $\mathbb{SD}$-S2 & \textbf{581}  & 501  & 192  & 106  \\ \hline
    \end{tabular}
  \end{center}
  \vspace{0mm}
\end{table}

%% file: tables/results_pedf.tex
\begin{table}[t]
  \begin{center}
  \scriptsize
    %\vspace{-3mm}
    \caption{\small{Total number of schedulable task sets in each scenario using different algorithms under P-EDF. In \textbf{Bold} the results of the best-performing algorithm.}}
    \label{tab:results_pedf}
     \renewcommand{\arraystretch}{1.2}
    \begin{tabular}{|c||c|c|c|}
    \hline
    \textbf{Scenario} & \emph{COMP} & \emph{CASE} & \emph{CaM}  \\ \hline \hline
    $\mathbb{AR}$-I $+$ $\mathbb{SH}$ $+$ $\mathbb{SD}$-B & 2096 & \textbf{2099} & 2071 \\ \hline
    $\mathbb{AR}$-I $+$ $\mathbb{SH}$ $+$ $\mathbb{SD}$-S1  & \textbf{1695} & 1692 & 1663 \\ \hline
    $\mathbb{AR}$-I $+$ $\mathbb{SH}$ $+$ $\mathbb{SD}$-S2  & 1447 & \textbf{1459} & 1397 \\ \hline
    $\mathbb{AR}$-I $+$ $\mathbb{WD}$ $+$ $\mathbb{SD}$-B   & \textbf{2113} & 2107 & 2074 \\ \hline
    $\mathbb{AR}$-I $+$ $\mathbb{WD}$ $+$ $\mathbb{SD}$-S1  & \textbf{1710} & 1699 & 1675 \\ \hline
    $\mathbb{AR}$-I $+$ $\mathbb{WD}$ $+$ $\mathbb{SD}$-S2  & 1424 & \textbf{1442} & 1383 \\ \hline
    $\mathbb{AR}$-II $+$ $\mathbb{SH}$ $+$ $\mathbb{SD}$-B  & 2522 & \textbf{2711} & 2617 \\ \hline
    $\mathbb{AR}$-II $+$ $\mathbb{SH}$ $+$ $\mathbb{SD}$-S1 & 923  & \textbf{977}  & 763  \\ \hline
    $\mathbb{AR}$-II $+$ $\mathbb{SH}$ $+$ $\mathbb{SD}$-S2 & 675  & \textbf{770}  & 433  \\ \hline
    $\mathbb{AR}$-II $+$ $\mathbb{WD}$ $+$ $\mathbb{SD}$-B  & 2519 & \textbf{2660} & 2578 \\ \hline
    $\mathbb{AR}$-II $+$ $\mathbb{WD}$ $+$ $\mathbb{SD}$-S1 & 931  & \textbf{1009} & 773  \\ \hline
    $\mathbb{AR}$-II $+$ $\mathbb{WD}$ $+$ $\mathbb{SD}$-S2 & 641  & \textbf{738}  & 406  \\ \hline
    \end{tabular}
  \end{center}
\end{table}

%% file: tables/rumtime.tex
% \usepackage{multirow}
\begin{table}[t]
\begin{center}
\scriptsize
\caption{\small{Running time comparison (avg/max) (in sec.).}}
%The average and maximum running time are reported as $\text{avg}/{\text{max}}$.}}
\label{tab:runtime}
\renewcommand{\arraystretch}{1.2}
\begin{tabular}{|c|l||l|l|l|l|l|}
\hline
\textbf{Scenario}                             & $n_p$    & \textit{COMP} & \textit{CASE} & \textit{IA$^3$} & \textit{PDPA} & \textit{CaM} \\ \hline\hline
\multicolumn{1}{|l|}{\multirow{2}{*}{NP-FP}}  & $16$ & $0.6/{2.2}$   & $0.5/{2.1}$  & $1.2/{3.7}$     &  $0.4/{3.9}$  & $9.4/{59}$     \\ \cline{2-7} 
\multicolumn{1}{|l|}{}                        & $32$ & $2.0/9.6$     & $1.5/7.4$    &  $1.8/7.7$      &  $0.8/8.7$    & $22/135$         \\ \hline
\multicolumn{1}{|l|}{\multirow{2}{*}{NP-EDF}} & $16$ & $4.7/44$      & $6.8/39$     &  $2.5/28$       &  $0.3/16$     & -            \\ \cline{2-7} 
\multicolumn{1}{|l|}{}                        & $32$ & $16/190$      & $19/168$     &  $4.0/68$       &  $0.5/18$     & -            \\ \hline
\multicolumn{1}{|l|}{\multirow{2}{*}{P-EDF}}  & $16$ & $0.2/1.0$     & $0.2/0.8$    &  -              & -             & $2.9/13$        \\ \cline{2-7} 
\multicolumn{1}{|l|}{}                        & $32$ & $0.5/2.4$     & $0.5/2.3$    &  -              & -             & $12/49$         \\ \hline
\end{tabular}
\end{center}
\vspace{0mm}
\end{table}

%% file: sections/conclusion.tex
\section{Conclusion and Future Work}
\label{sec:concl}
Unlike in preemptive scheduling, a lower-priority non-preemptive task can \emph{block} a higher-priority task, significantly impacting the schedulability of a task set.
Also, intuitively, clustering tasks with similar \emph{cache sensitivities} can maximize the cache utilization and improve the schedulability.
This paper provides useful insights on the \emph{trade-offs} between blocking and cache sensitivity in establishing the schedulability of NP-FP tasks on multi-core processors.

We propose a \emph{multi-layer hybrid design space exploration framework} to solve the joint problem of cache partitioning and task allocation. Our extensive experimental evaluation against state-of-the-art algorithms shows that our framework can considerably improve real-time schedulability even when cache sensitivities of tasks cannot be fully exploited.
Although some optimization strategies of the framework are specifically designed for NP-FP tasks, we show that the framework also achieves good schedulability results for \emph{preemptive} and \emph{non-preemptive EDF} tasks.

While this paper evaluates blocking and cache sensitivity separately, we will investigate in the future how to combine these two factors into one metric that can drive task allocation.
Besides, we can possibly determine sets of mutually compatible tasks and then, for each set, perform task allocation based on cache sensitivity.
Naturally, we will also consider complex task models (\eg, directed acyclic graphs) and platform settings (\eg, memory bandwidth regulation \cite{yun2013memguard}).

%% file: appendix.tex
\appendix

The appendix includes:
\begin{enumerate}
    \item the complete visualization of schedulability comparison results under NP-FP (Figure~\ref{fig:npfp}), NP-EDF (Figure~\ref{fig:npedf}), and P-EDF (Figure~\ref{fig:pedf}) scheduling policies;
    \item the complete set of benchmark profiles (Figure~7 - 56).
\end{enumerate}

\input{figures/sched_results_complete_npfp}

\input{figures/sched_results_complete_npedf}

\input{figures/sched_results_complete_pedf}

\input{figures/bench_plots}

\clearpage

%% file: figures/sched_results_complete_npfp.tex
\begin{figure*}[b]
    \centering
    \begin{subfigure}{.33\textwidth}
      \centering
      \includegraphics[width=0.95\linewidth]{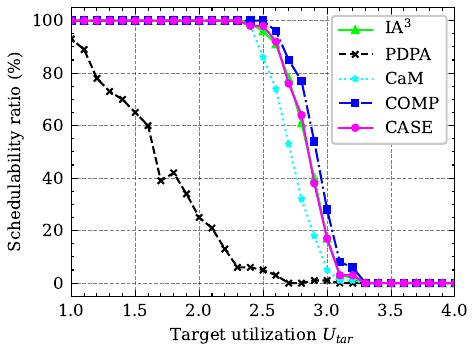}
      \caption{$\mathbb{AR}$-I $+ \mathbb{SH} + \mathbb{SD}$-B.}
      \label{fig:ar1_sh_b}
    \end{subfigure}%
    \begin{subfigure}{.33\textwidth}
      \centering
      \includegraphics[width=0.95\linewidth]{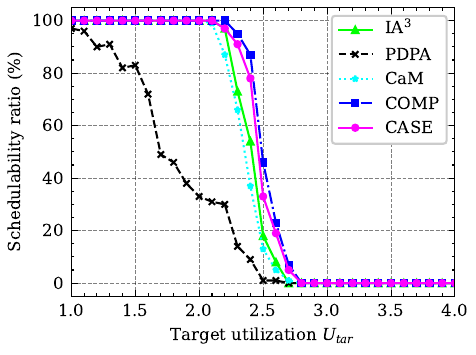}
      \caption{$\mathbb{AR}$-I $+ \mathbb{SH} + \mathbb{SD}$-S1.}
      \label{fig:ar1_sh_s1}
    \end{subfigure}%
    \begin{subfigure}{.33\textwidth}
      \centering
      \includegraphics[width=0.95\linewidth]{figures/npfp/short_period_synthetic_30_1024_16.pdf}
      \caption{$\mathbb{AR}$-I $+ \mathbb{SH} + \mathbb{SD}$-S2.}
      \label{fig:ar1_sh_s2}
    \end{subfigure}%
    \\
    \begin{subfigure}{.33\textwidth}
      \centering
      \includegraphics[width=0.95\linewidth]{figures/npfp/wide_period_benchmark_1024_16.pdf}
      \caption{$\mathbb{AR}$-I $+ \mathbb{WD} + \mathbb{SD}$-B.}
      \label{fig:ar1_wd_b}
    \end{subfigure}%
    \begin{subfigure}{.33\textwidth}
      \centering
      \includegraphics[width=0.95\linewidth]{figures/npfp/wide_period_synthetic_24_1024_16.pdf}
      \caption{$\mathbb{AR}$-I $+ \mathbb{WD} + \mathbb{SD}$-S1.}
      \label{fig:ar1_wd_s1}
    \end{subfigure}%
    \begin{subfigure}{.33\textwidth}
      \centering
      \includegraphics[width=0.95\linewidth]{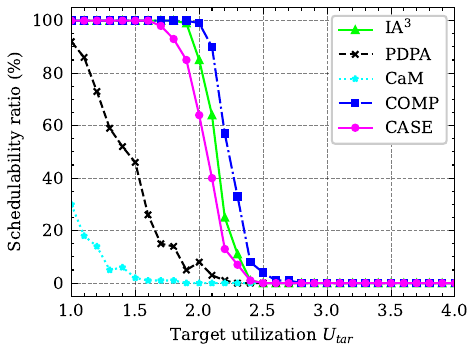}
      \caption{$\mathbb{AR}$-I $+ \mathbb{WD} + \mathbb{SD}$-S2.}
      \label{fig:ar1_wd_s2}
    \end{subfigure}%
    \\
    \begin{subfigure}{.33\textwidth}
      \centering
      \includegraphics[width=0.95\linewidth]{figures/npfp/short_period_benchmark_2048_32.pdf}
      \caption{$\mathbb{AR}$-II $+ \mathbb{SH} + \mathbb{SD}$-B.}
      \label{fig:ar2_sh_b}
    \end{subfigure}%
    \begin{subfigure}{.33\textwidth}
      \centering
      \includegraphics[width=0.95\linewidth]{figures/npfp/short_period_synthetic_60_2048_32.pdf}
      \caption{$\mathbb{AR}$-II $+ \mathbb{SH} + \mathbb{SD}$-S1.}
      \label{fig:ar2_sh_s1}
    \end{subfigure}%
    \begin{subfigure}{.33\textwidth}
      \centering
      \includegraphics[width=0.95\linewidth]{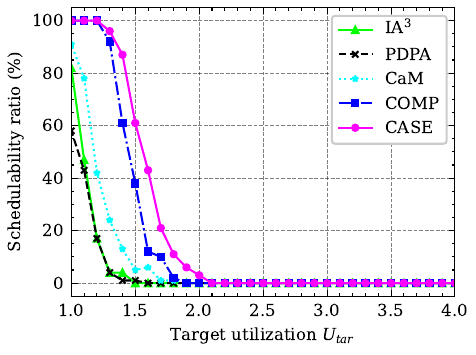}
      \caption{$\mathbb{AR}$-II $+ \mathbb{SH} + \mathbb{SD}$-S2.}
      \label{fig:ar2_sh_s2}
    \end{subfigure}%
    \\
    \begin{subfigure}{.33\textwidth}
      \centering
      \includegraphics[width=0.95\linewidth]{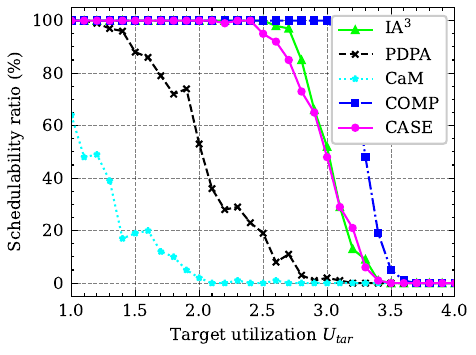}
      \caption{$\mathbb{AR}$-II $+ \mathbb{WD} + \mathbb{SD}$-B.}
      \label{fig:ar2_wd_b}
    \end{subfigure}%
    \begin{subfigure}{.33\textwidth}
      \centering
      \includegraphics[width=0.95\linewidth]{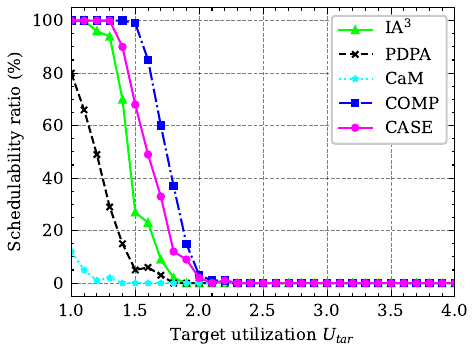}
      \caption{$\mathbb{AR}$-II $+ \mathbb{WD} + \mathbb{SD}$-S1.}
      \label{fig:ar2_wd_s1}
    \end{subfigure}%
    \begin{subfigure}{.33\textwidth}
      \centering
      \includegraphics[width=0.95\linewidth]{figures/npfp/wide_period_synthetic_100_2048_32.pdf}
      \caption{$\mathbb{AR}$-II $+ \mathbb{WD} + \mathbb{SD}$-S2.}
      \label{fig:ar2_wd_s2}
    \end{subfigure}%
    \caption{Schedulability ratio (\textit{i.e.}, $\frac{\text{\#schedulable task sets}}{\text{\#total task sets}} \times 100$ \%) of proposed schemes and baseline algorithms under NP-FP.}
    \label{fig:npfp}
\end{figure*}

%% file: figures/sched_results_complete_npedf.tex
\begin{figure*}[ht]
    \centering
    \begin{subfigure}{.33\textwidth}
      \centering
      \includegraphics[width=0.95\linewidth]{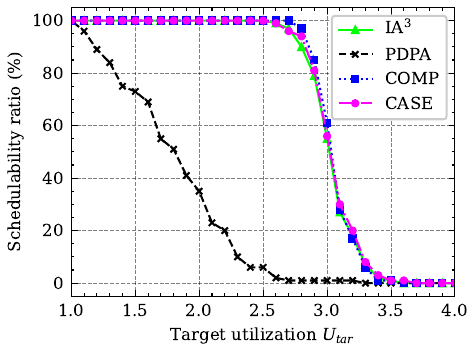}
      \caption{$\mathbb{AR}$-I $+ \mathbb{SH} + \mathbb{SD}$-B.}
      \label{fig:ar1_sh_b}
    \end{subfigure}%
    \begin{subfigure}{.33\textwidth}
      \centering
      \includegraphics[width=0.95\linewidth]{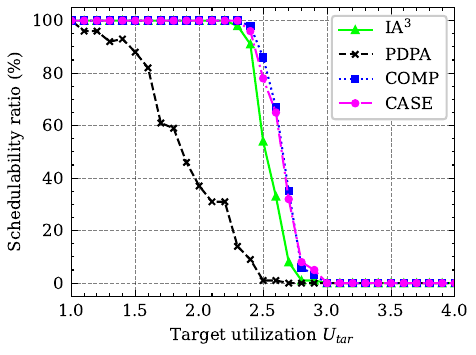}
      \caption{$\mathbb{AR}$-I $+ \mathbb{SH} + \mathbb{SD}$-S1.}
      \label{fig:ar1_sh_s1}
    \end{subfigure}%
    \begin{subfigure}{.33\textwidth}
      \centering
      \includegraphics[width=0.95\linewidth]{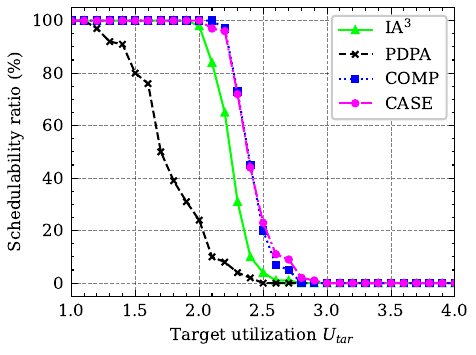}
      \caption{$\mathbb{AR}$-I $+ \mathbb{SH} + \mathbb{SD}$-S2.}
      \label{fig:ar1_sh_s2}
    \end{subfigure}%
    \\
    \begin{subfigure}{.33\textwidth}
      \centering
      \includegraphics[width=0.95\linewidth]{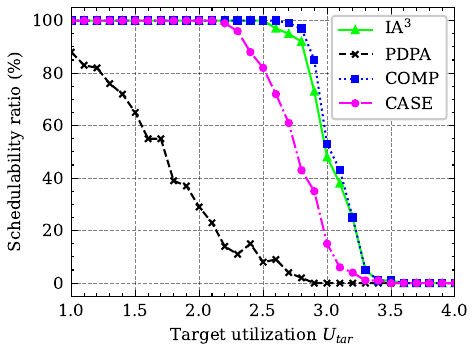}
      \caption{$\mathbb{AR}$-I $+ \mathbb{WD} + \mathbb{SD}$-B.}
      \label{fig:ar1_wd_b}
    \end{subfigure}%
    \begin{subfigure}{.33\textwidth}
      \centering
      \includegraphics[width=0.95\linewidth]{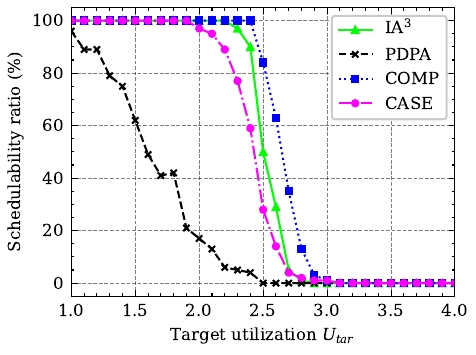}
      \caption{$\mathbb{AR}$-I $+ \mathbb{WD} + \mathbb{SD}$-S1.}
      \label{fig:ar1_wd_s1}
    \end{subfigure}%
    \begin{subfigure}{.33\textwidth}
      \centering
      \includegraphics[width=0.95\linewidth]{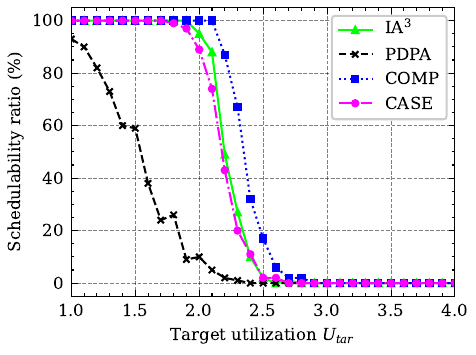}
      \caption{$\mathbb{AR}$-I $+ \mathbb{WD} + \mathbb{SD}$-S2.}
      \label{fig:ar1_wd_s2}
    \end{subfigure}%
    \\
    \begin{subfigure}{.33\textwidth}
      \centering
      \includegraphics[width=0.95\linewidth]{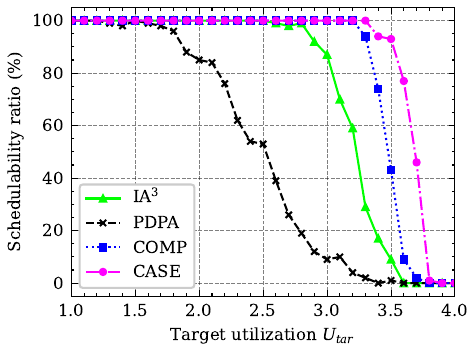}
      \caption{$\mathbb{AR}$-II $+ \mathbb{SH} + \mathbb{SD}$-B.}
      \label{fig:ar2_sh_b}
    \end{subfigure}%
    \begin{subfigure}{.33\textwidth}
      \centering
      \includegraphics[width=0.95\linewidth]{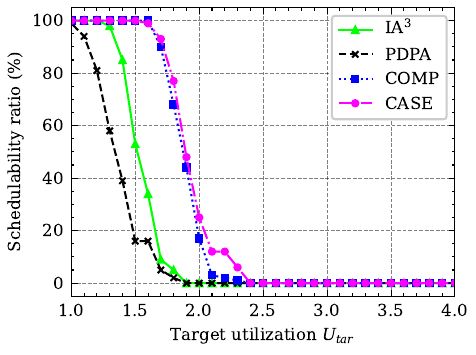}
      \caption{$\mathbb{AR}$-II $+ \mathbb{SH} + \mathbb{SD}$-S1.}
      \label{fig:ar2_sh_s1}
    \end{subfigure}%
    \begin{subfigure}{.33\textwidth}
      \centering
      \includegraphics[width=0.95\linewidth]{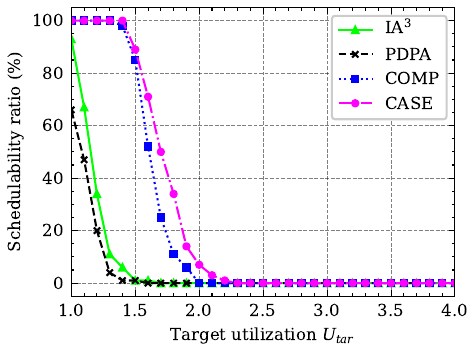}
      \caption{$\mathbb{AR}$-II $+ \mathbb{SH} + \mathbb{SD}$-S2.}
      \label{fig:ar2_sh_s2}
    \end{subfigure}%
    \\
    \begin{subfigure}{.33\textwidth}
      \centering
      \includegraphics[width=0.95\linewidth]{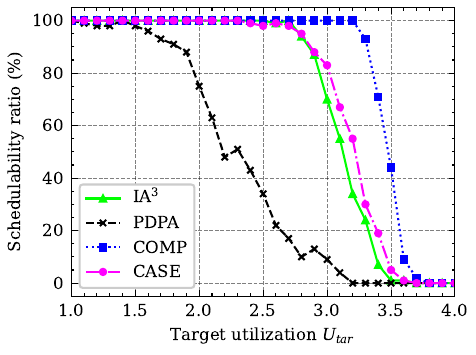}
      \caption{$\mathbb{AR}$-II $+ \mathbb{WD} + \mathbb{SD}$-B.}
      \label{fig:ar2_wd_b}
    \end{subfigure}%
    \begin{subfigure}{.33\textwidth}
      \centering
      \includegraphics[width=0.95\linewidth]{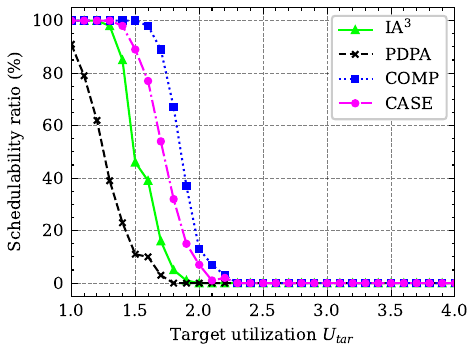}
      \caption{$\mathbb{AR}$-II $+ \mathbb{WD} + \mathbb{SD}$-S1.}
      \label{fig:ar2_wd_s1}
    \end{subfigure}%
    \begin{subfigure}{.33\textwidth}
      \centering
      \includegraphics[width=0.95\linewidth]{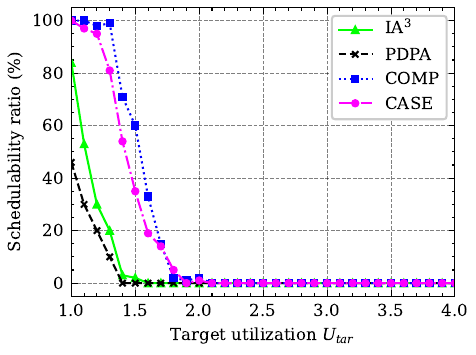}
      \caption{$\mathbb{AR}$-II $+ \mathbb{WD} + \mathbb{SD}$-S2.}
      \label{fig:ar2_wd_s2}
    \end{subfigure}%
    \caption{Schedulability ratio (\textit{i.e.}, $\frac{\text{\#schedulable task sets}}{\text{\#total task sets}} \times 100$ \%) of proposed schemes and baseline algorithms under NP-EDF.}
    \label{fig:npedf}
\end{figure*}

%% file: figures/sched_results_complete_pedf.tex
\begin{figure*}[ht]
    \centering
    \begin{subfigure}{.33\textwidth}
      \centering
      \includegraphics[width=0.95\linewidth]{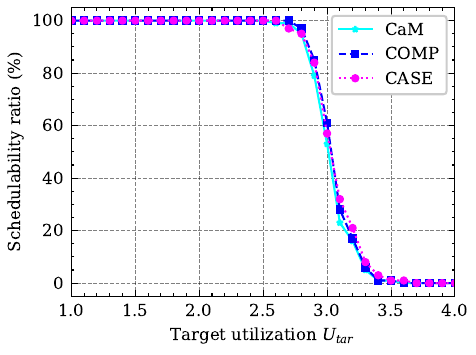}
      \caption{$\mathbb{AR}$-I $+ \mathbb{SH} + \mathbb{SD}$-B.}
      \label{fig:ar1_sh_b}
    \end{subfigure}%
    \begin{subfigure}{.33\textwidth}
      \centering
      \includegraphics[width=0.95\linewidth]{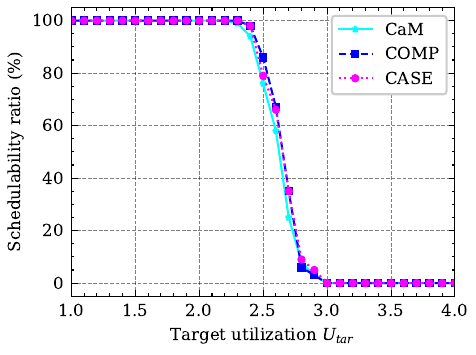}
      \caption{$\mathbb{AR}$-I $+ \mathbb{SH} + \mathbb{SD}$-S1.}
      \label{fig:ar1_sh_s1}
    \end{subfigure}%
    \begin{subfigure}{.33\textwidth}
      \centering
      \includegraphics[width=0.95\linewidth]{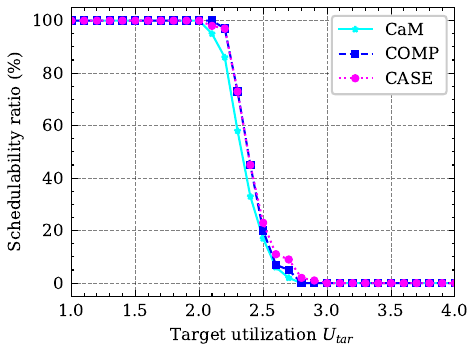}
      \caption{$\mathbb{AR}$-I $+ \mathbb{SH} + \mathbb{SD}$-S2.}
      \label{fig:ar1_sh_s2}
    \end{subfigure}%
    \\
    \begin{subfigure}{.33\textwidth}
      \centering
      \includegraphics[width=0.95\linewidth]{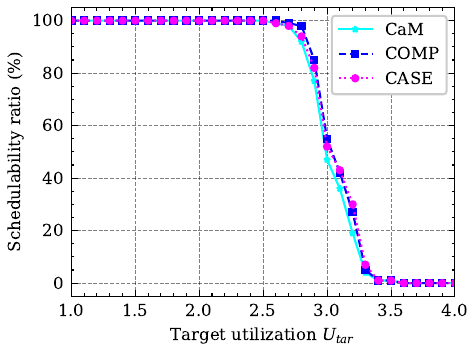}
      \caption{$\mathbb{AR}$-I $+ \mathbb{WD} + \mathbb{SD}$-B.}
      \label{fig:ar1_wd_b}
    \end{subfigure}%
    \begin{subfigure}{.33\textwidth}
      \centering
      \includegraphics[width=0.95\linewidth]{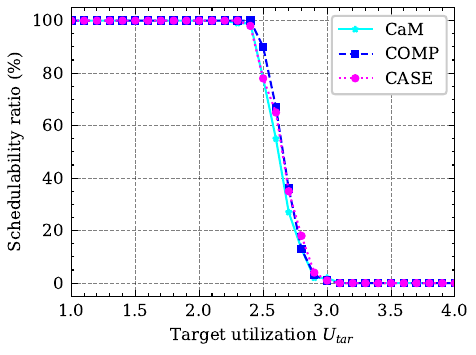}
      \caption{$\mathbb{AR}$-I $+ \mathbb{WD} + \mathbb{SD}$-S1.}
      \label{fig:ar1_wd_s1}
    \end{subfigure}%
    \begin{subfigure}{.33\textwidth}
      \centering
      \includegraphics[width=0.95\linewidth]{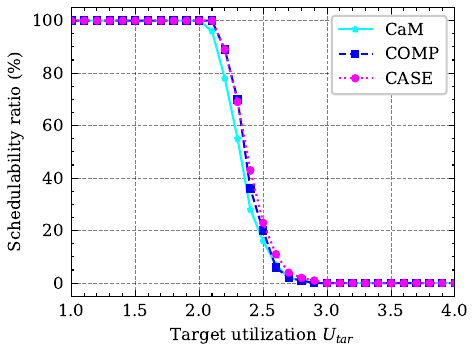}
      \caption{$\mathbb{AR}$-I $+ \mathbb{WD} + \mathbb{SD}$-S2.}
      \label{fig:ar1_wd_s2}
    \end{subfigure}%
    \\
    \begin{subfigure}{.33\textwidth}
      \centering
      \includegraphics[width=0.95\linewidth]{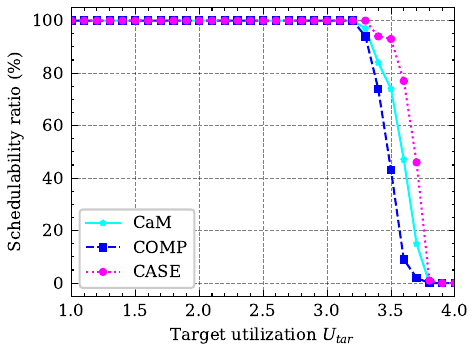}
      \caption{$\mathbb{AR}$-II $+ \mathbb{SH} + \mathbb{SD}$-B.}
      \label{fig:ar2_sh_b}
    \end{subfigure}%
    \begin{subfigure}{.33\textwidth}
      \centering
      \includegraphics[width=0.95\linewidth]{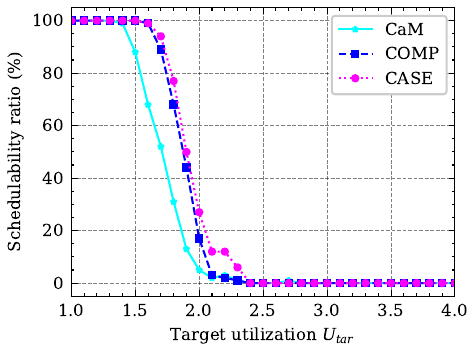}
      \caption{$\mathbb{AR}$-II $+ \mathbb{SH} + \mathbb{SD}$-S1.}
      \label{fig:ar2_sh_s1}
    \end{subfigure}%
    \begin{subfigure}{.33\textwidth}
      \centering
      \includegraphics[width=0.95\linewidth]{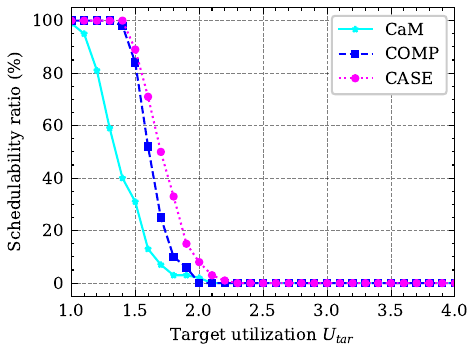}
      \caption{$\mathbb{AR}$-II $+ \mathbb{SH} + \mathbb{SD}$-S2.}
      \label{fig:ar2_sh_s2}
    \end{subfigure}%
    \\
    \begin{subfigure}{.33\textwidth}
      \centering
      \includegraphics[width=0.95\linewidth]{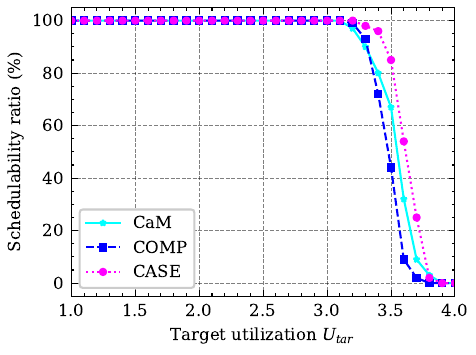}
      \caption{$\mathbb{AR}$-II $+ \mathbb{WD} + \mathbb{SD}$-B.}
      \label{fig:ar2_wd_b}
    \end{subfigure}%
    \begin{subfigure}{.33\textwidth}
      \centering
      \includegraphics[width=0.95\linewidth]{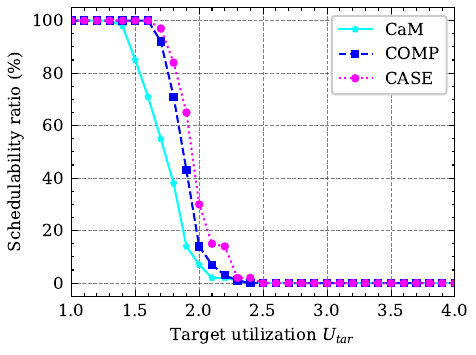}
      \caption{$\mathbb{AR}$-II $+ \mathbb{WD} + \mathbb{SD}$-S1.}
      \label{fig:ar2_wd_s1}
    \end{subfigure}%
    \begin{subfigure}{.33\textwidth}
      \centering
      \includegraphics[width=0.95\linewidth]{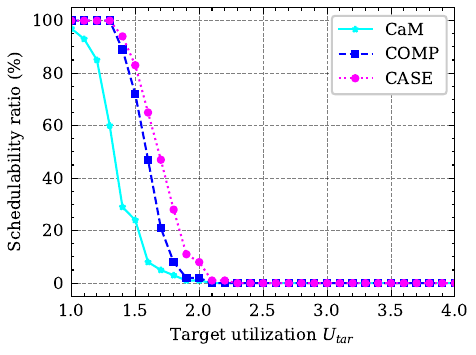}
      \caption{$\mathbb{AR}$-II $+ \mathbb{WD} + \mathbb{SD}$-S2.}
      \label{fig:ar2_wd_s2}
    \end{subfigure}%
    \caption{Schedulability ratio (\textit{i.e.}, $\frac{\text{\#schedulable task sets}}{\text{\#total task sets}} \times 100$ \%) of proposed schemes and baseline algorithms under P-EDF.}
    \label{fig:pedf}
\end{figure*}

%% file: figures/bench_plots.tex
\begin{figure}[t]
\centering
\includegraphics[width=0.9\columnwidth]{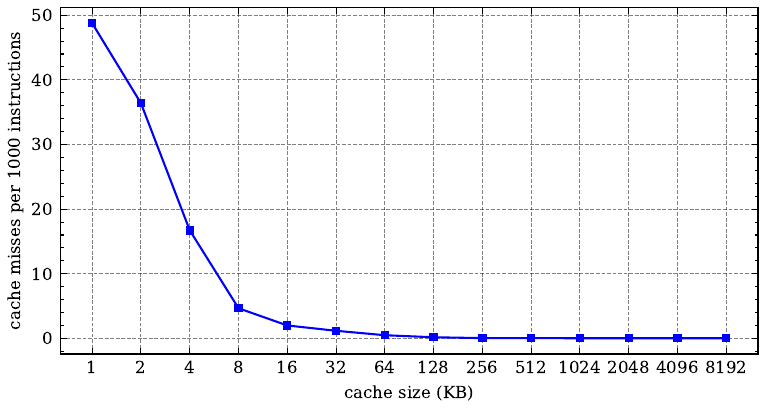}
\caption{stats svm cif.log}
\end{figure}

\begin{figure}[t]
\centering
\includegraphics[width=0.9\columnwidth]{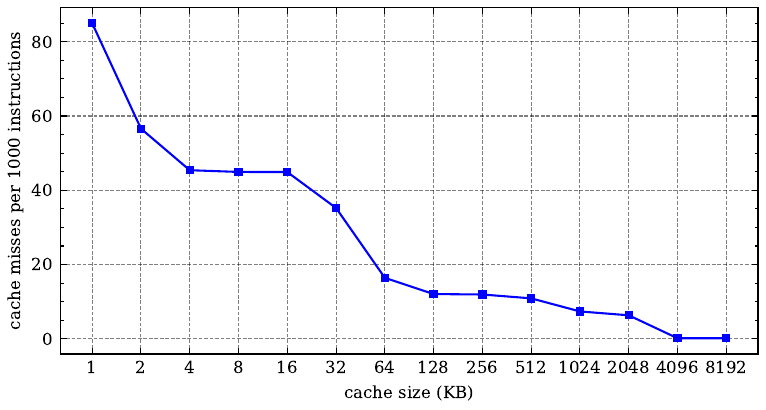}
\caption{stats disparity cif.log}
\end{figure}

\begin{figure}[t]
\centering
\includegraphics[width=0.9\columnwidth]{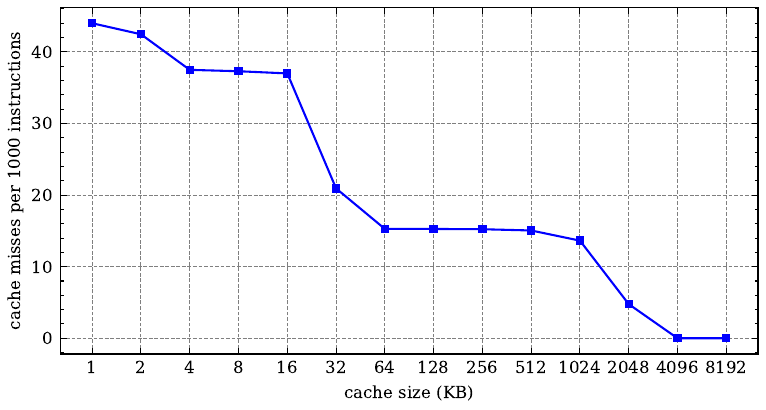}
\caption{stats rbm medium.log}
\end{figure}

\begin{figure}[t]
\centering
\includegraphics[width=0.9\columnwidth]{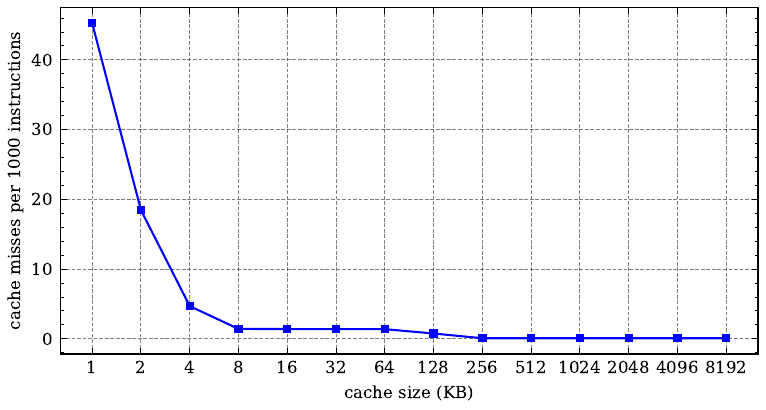}
\caption{stats kmeans small.log}
\end{figure}

\begin{figure}[t]
\centering
\includegraphics[width=0.9\columnwidth]{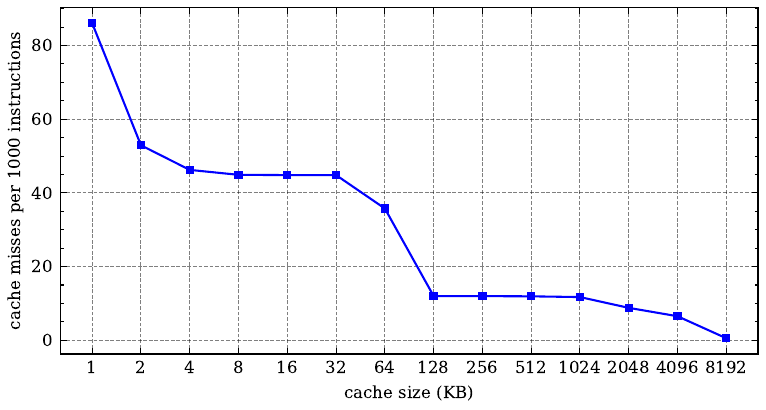}
\caption{stats disparity vga.log}
\end{figure}

\begin{figure}[t]
\centering
\includegraphics[width=0.9\columnwidth]{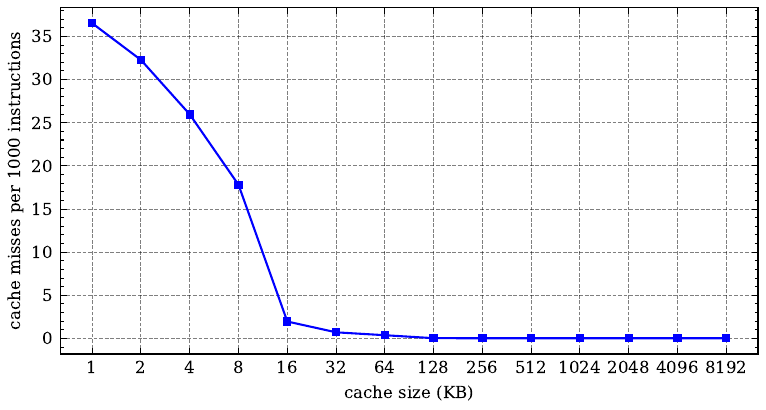}
\caption{stats svd3 small.log}
\end{figure}

\begin{figure}[t]
\centering
\includegraphics[width=0.9\columnwidth]{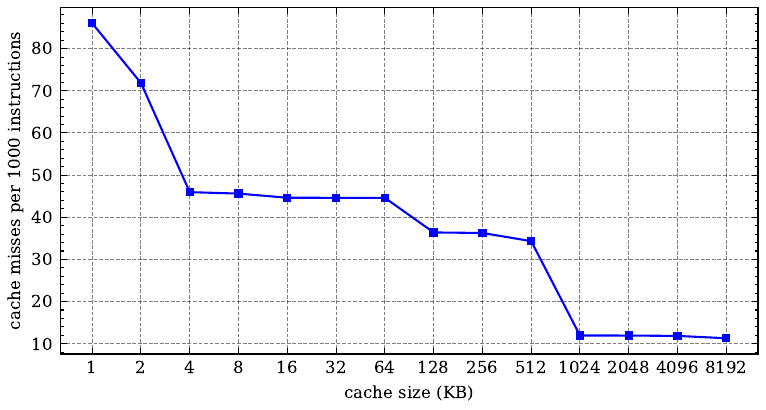}
\caption{stats disparity fullhd.log}
\end{figure}

\begin{figure}[t]
\centering
\includegraphics[width=0.9\columnwidth]{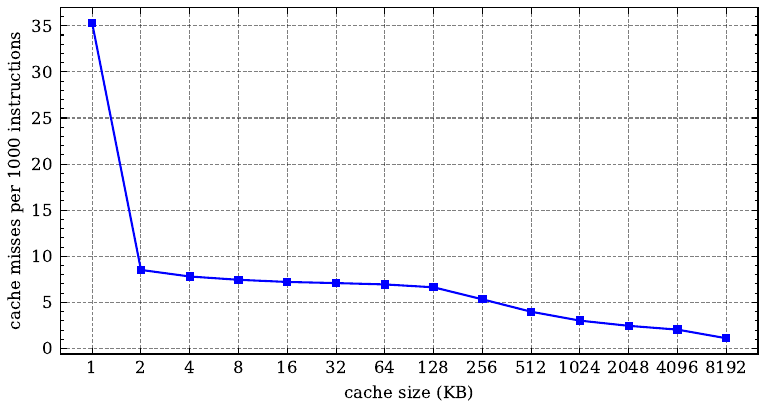}
\caption{stats mser vga.log}
\end{figure}

\begin{figure}[t]
\centering
\includegraphics[width=0.9\columnwidth]{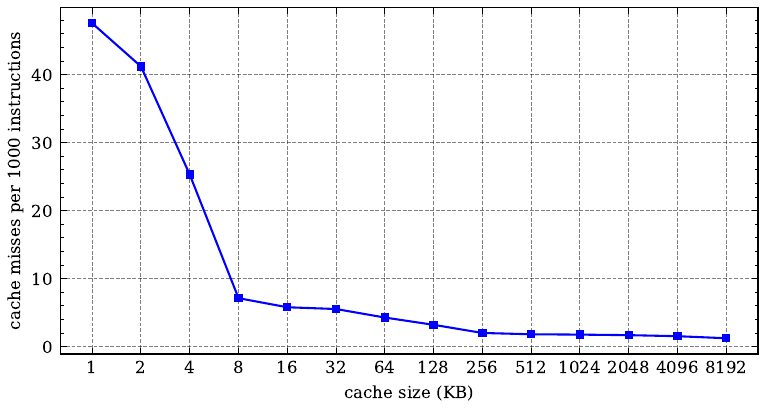}
\caption{stats sift vga.log}
\end{figure}

\begin{figure}[t]
\centering
\includegraphics[width=0.9\columnwidth]{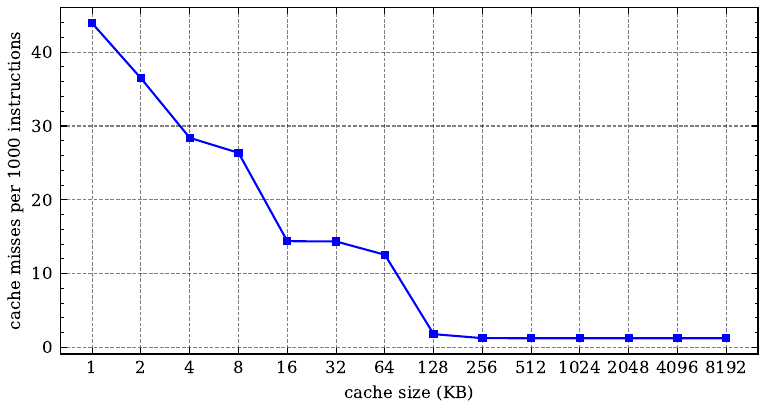}
\caption{stats kmeans large.log}
\end{figure}

\begin{figure}[t]
\centering
\includegraphics[width=0.9\columnwidth]{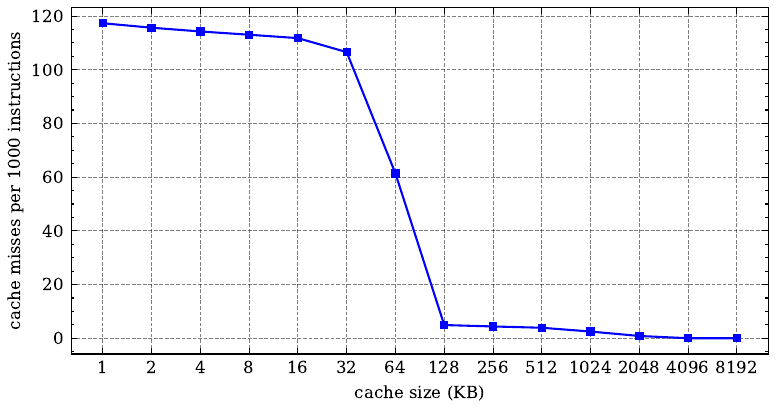}
\caption{stats pca medium.log}
\end{figure}

\begin{figure}[t]
\centering
\includegraphics[width=0.9\columnwidth]{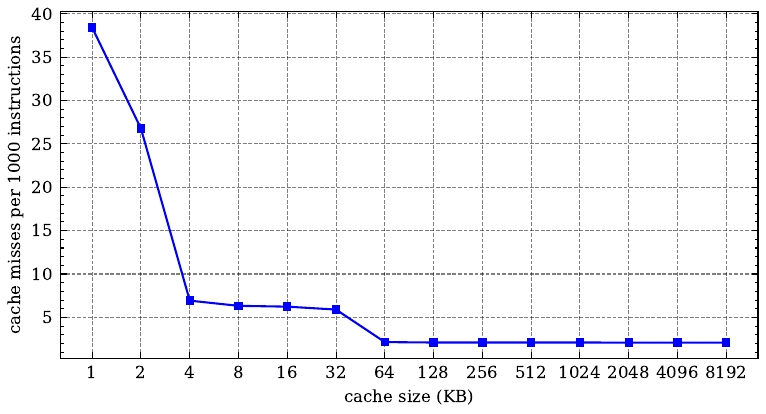}
\caption{stats liblinear medium.log}
\end{figure}

\begin{figure}[t]
\centering
\includegraphics[width=0.9\columnwidth]{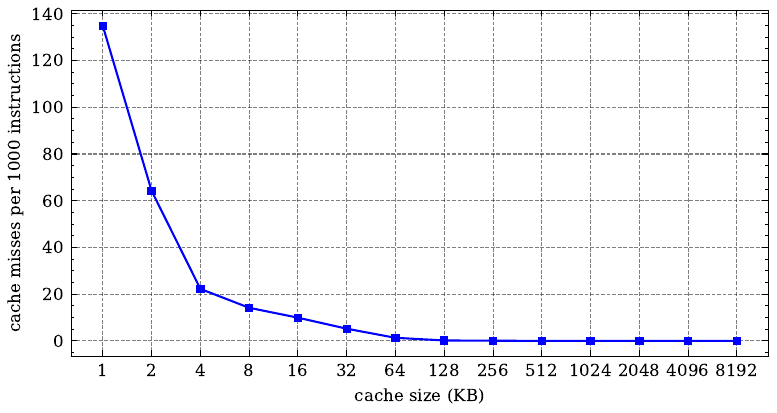}
\caption{stats lda small.log}
\end{figure}

\begin{figure}[t]
\centering
\includegraphics[width=0.9\columnwidth]{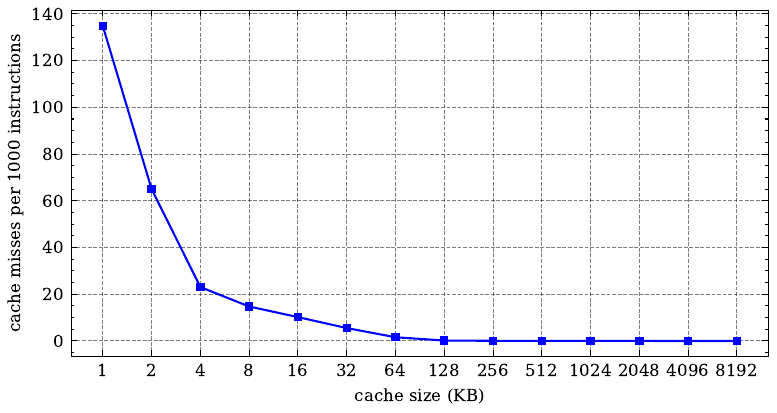}
\caption{stats lda large.log}
\end{figure}

\begin{figure}[t]
\centering
\includegraphics[width=0.9\columnwidth]{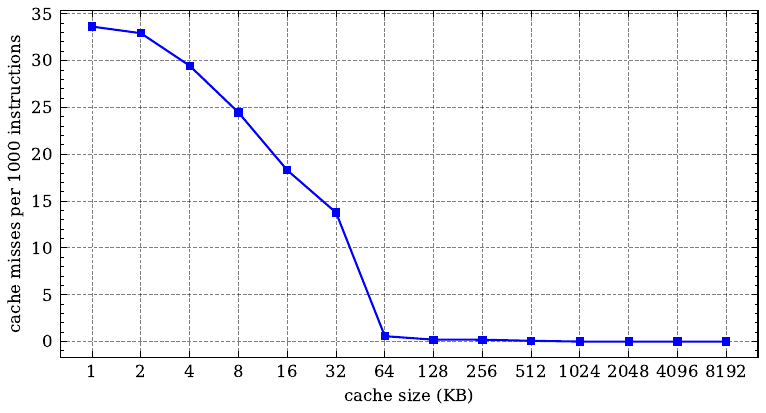}
\caption{stats srr small.log}
\end{figure}

\begin{figure}[t]
\centering
\includegraphics[width=0.9\columnwidth]{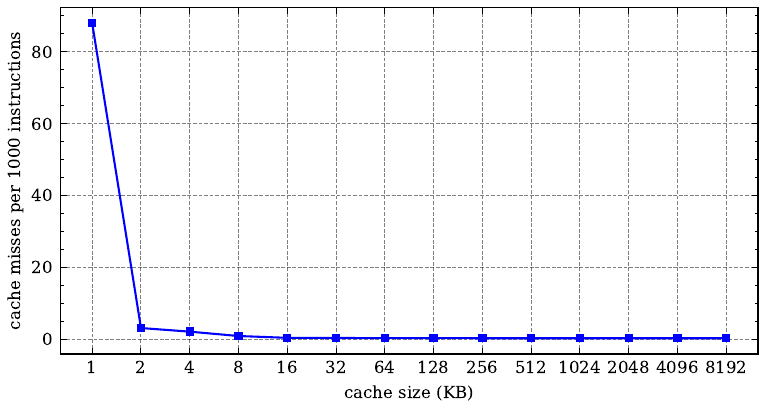}
\caption{stats sphinx small.log}
\end{figure}

\begin{figure}[t]
\centering
\includegraphics[width=0.9\columnwidth]{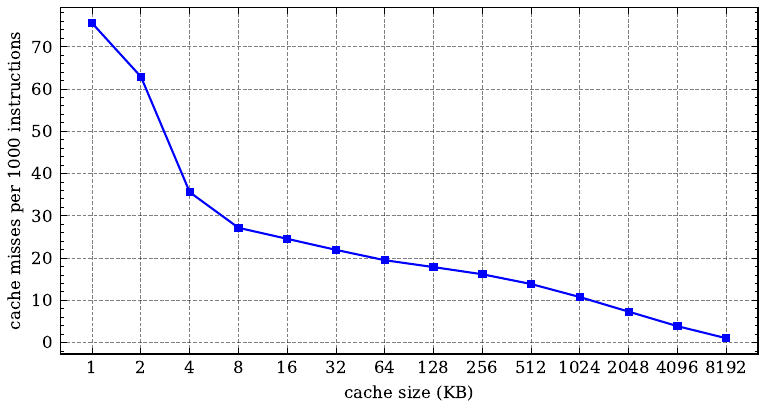}
\caption{stats sphinx large.log}
\end{figure}

\begin{figure}[t]
\centering
\includegraphics[width=0.9\columnwidth]{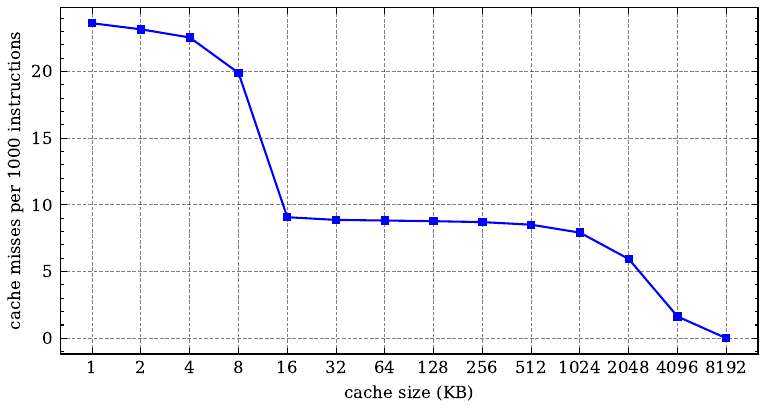}
\caption{stats spc medium.log}
\end{figure}

\begin{figure}[t]
\centering
\includegraphics[width=0.9\columnwidth]{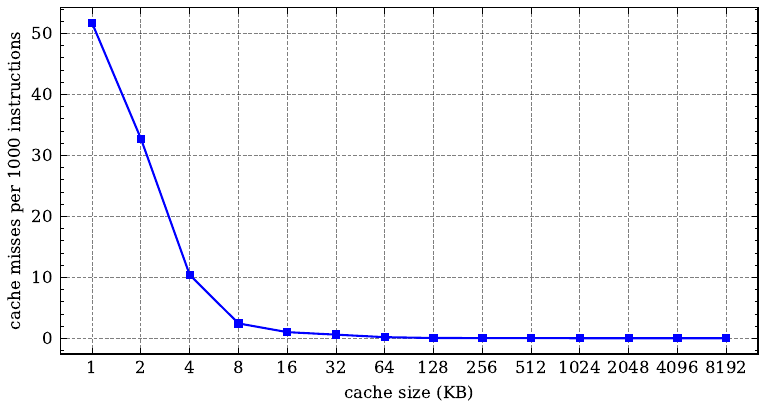}
\caption{stats svm sqcif.log}
\end{figure}

\begin{figure}[t]
\centering
\includegraphics[width=0.9\columnwidth]{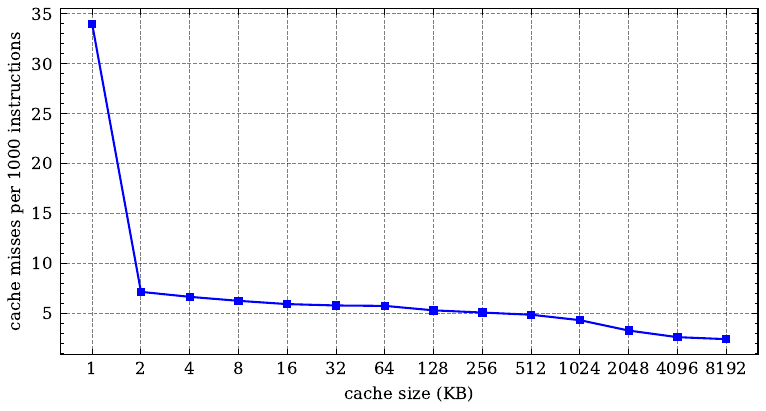}
\caption{stats mser fullhd.log}
\end{figure}

\begin{figure}[t]
\centering
\includegraphics[width=0.9\columnwidth]{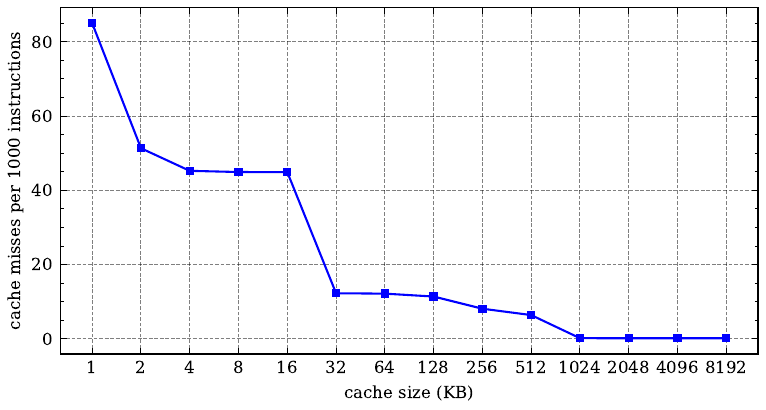}
\caption{stats disparity qcif.log}
\end{figure}

\begin{figure}[t]
\centering
\includegraphics[width=0.9\columnwidth]{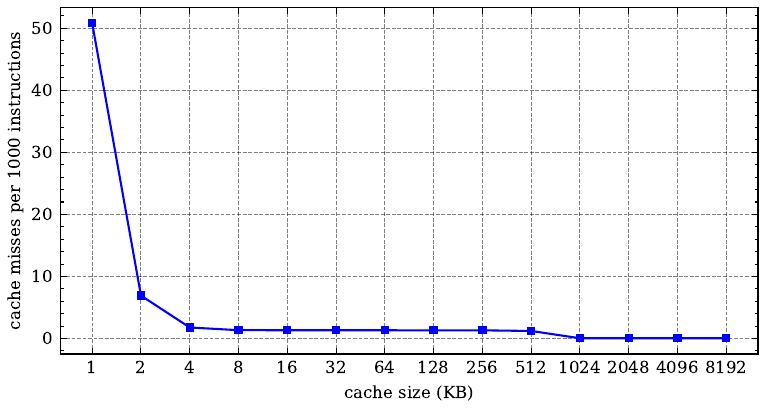}
\caption{stats kmeans medium.log}
\end{figure}

\begin{figure}[t]
\centering
\includegraphics[width=0.9\columnwidth]{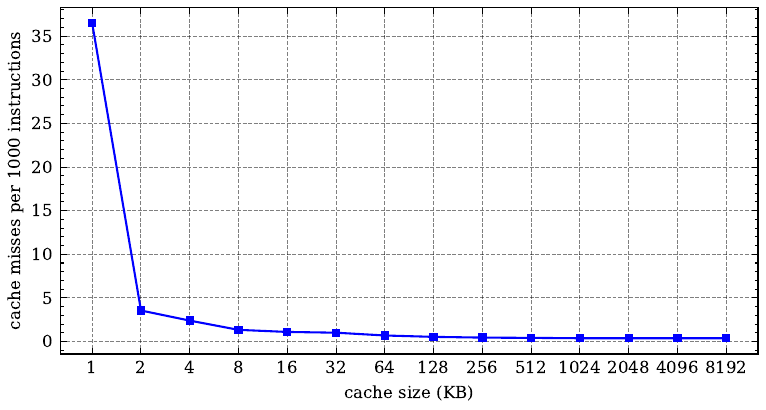}
\caption{stats me small.log}
\end{figure}

\begin{figure}[t]
\centering
\includegraphics[width=0.9\columnwidth]{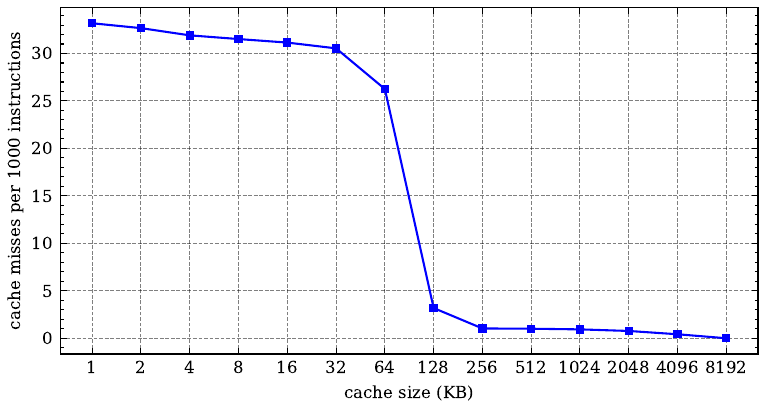}
\caption{stats svd3 large.log}
\end{figure}

\begin{figure}[t]
\centering
\includegraphics[width=0.9\columnwidth]{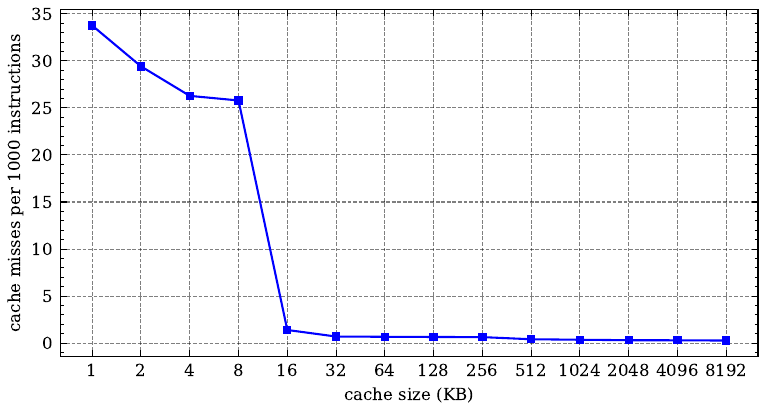}
\caption{stats me medium.log}
\end{figure}

\begin{figure}[t]
\centering
\includegraphics[width=0.9\columnwidth]{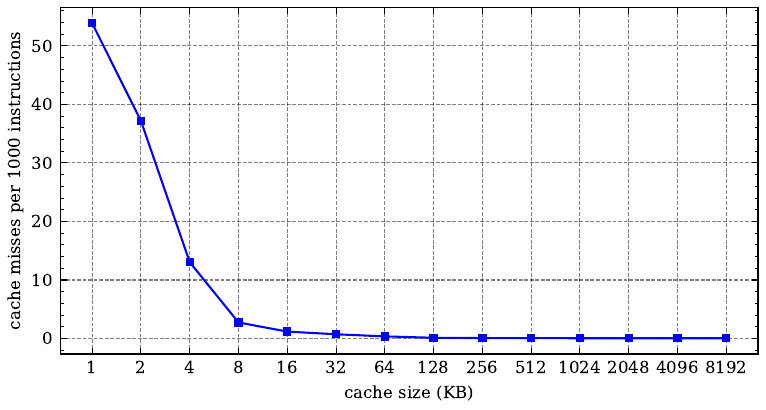}
\caption{stats svm qcif.log}
\end{figure}

\begin{figure}[t]
\centering
\includegraphics[width=0.9\columnwidth]{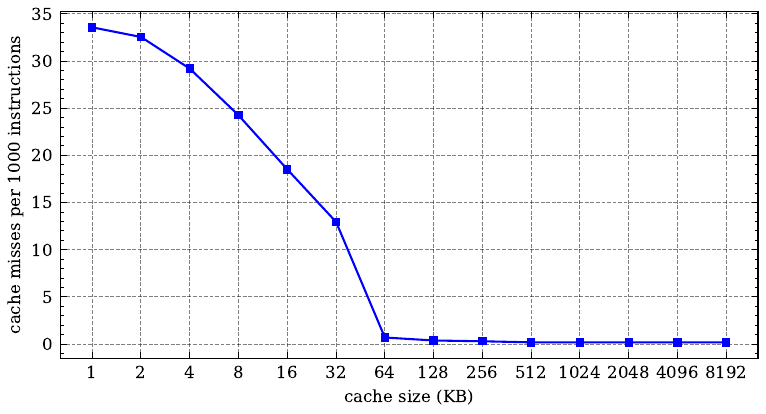}
\caption{stats srr large.log}
\end{figure}

\begin{figure}[t]
\centering
\includegraphics[width=0.9\columnwidth]{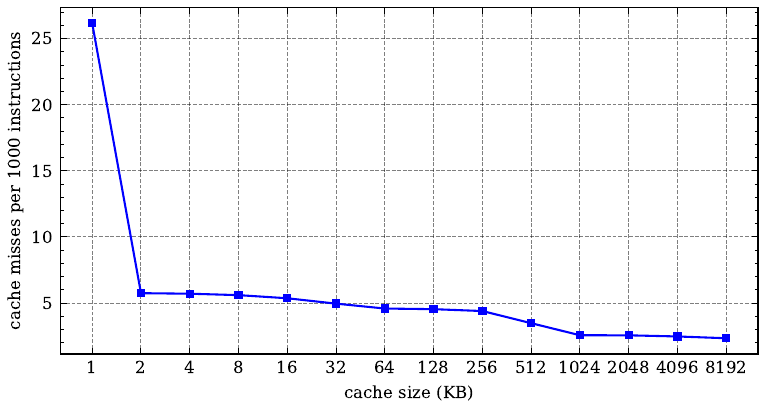}
\caption{stats tracking fullhd.log}
\end{figure}

\begin{figure}[t]
\centering
\includegraphics[width=0.9\columnwidth]{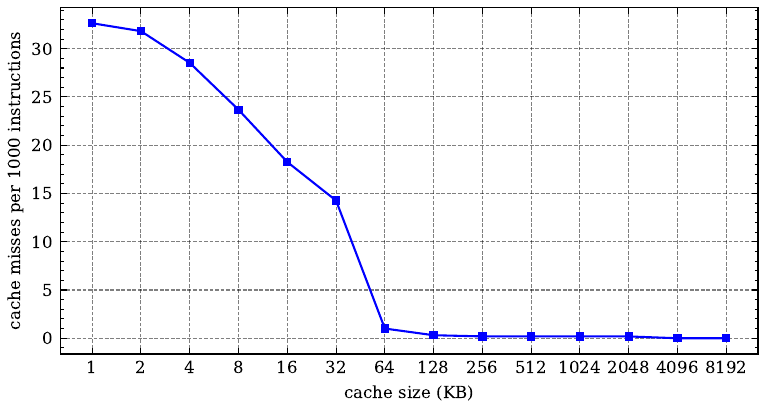}
\caption{stats srr medium.log}
\end{figure}

\begin{figure}[t]
\centering
\includegraphics[width=0.9\columnwidth]{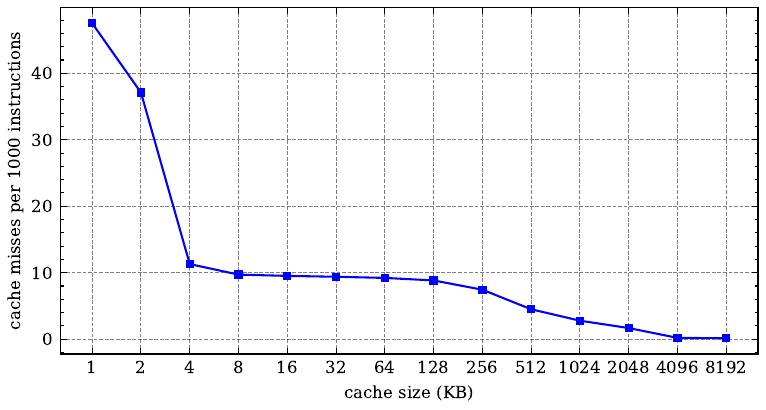}
\caption{stats liblinear small.log}
\end{figure}

\begin{figure}[t]
\centering
\includegraphics[width=0.9\columnwidth]{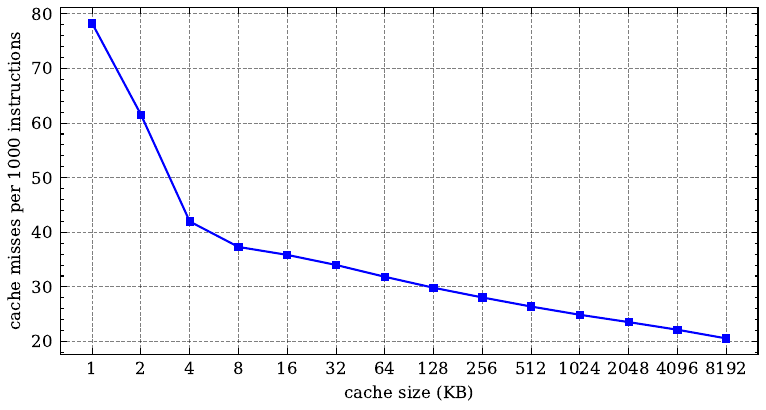}
\caption{stats liblinear large.log}
\end{figure}

\begin{figure}[t]
\centering
\includegraphics[width=0.9\columnwidth]{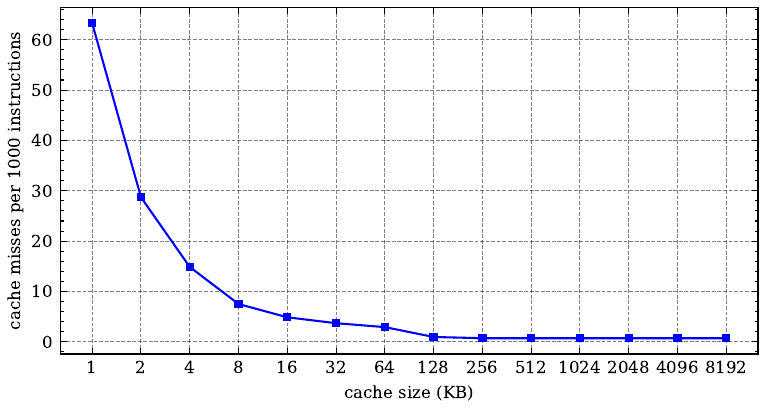}
\caption{stats disparity sim.log}
\end{figure}

\begin{figure}[t]
\centering
\includegraphics[width=0.9\columnwidth]{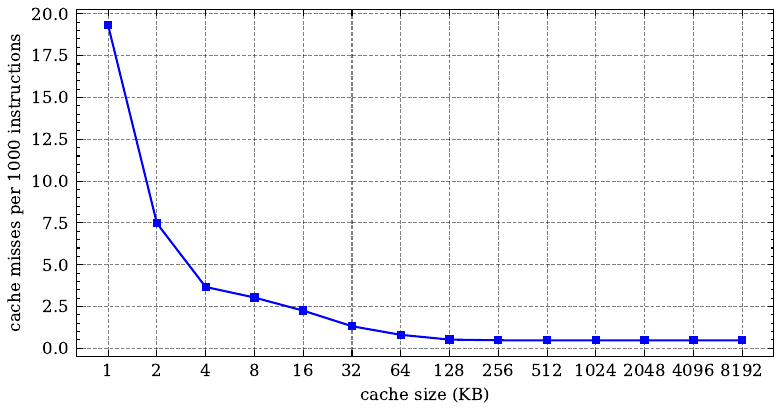}
\caption{stats sift sim.log}
\end{figure}

\begin{figure}[t]
\centering
\includegraphics[width=0.9\columnwidth]{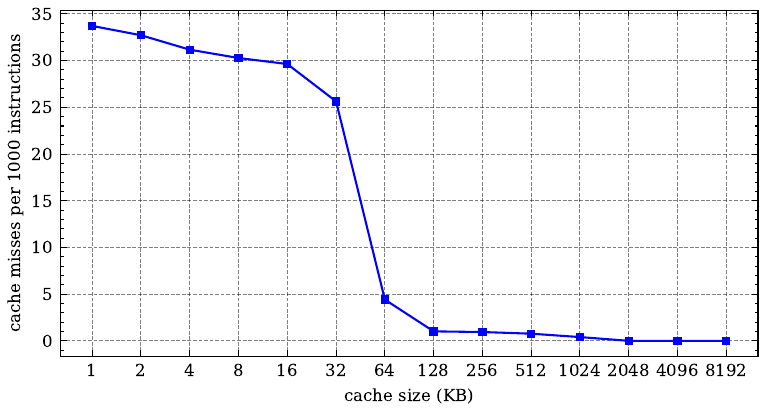}
\caption{stats svd3 medium.log}
\end{figure}

\begin{figure}[t]
\centering
\includegraphics[width=0.9\columnwidth]{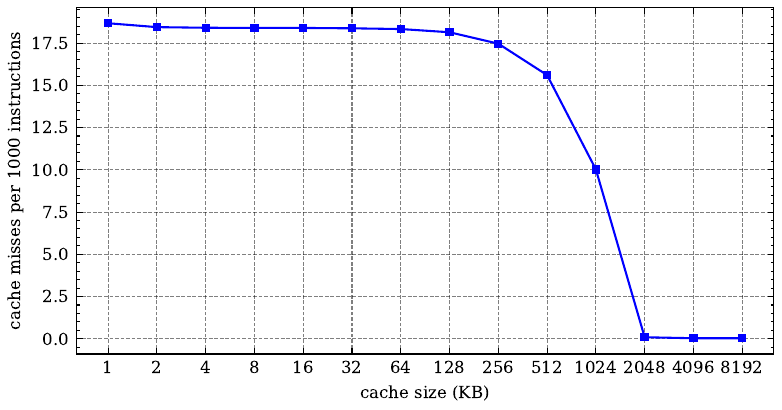}
\caption{stats stitch fullhd.log}
\end{figure}

\begin{figure}[t]
\centering
\includegraphics[width=0.9\columnwidth]{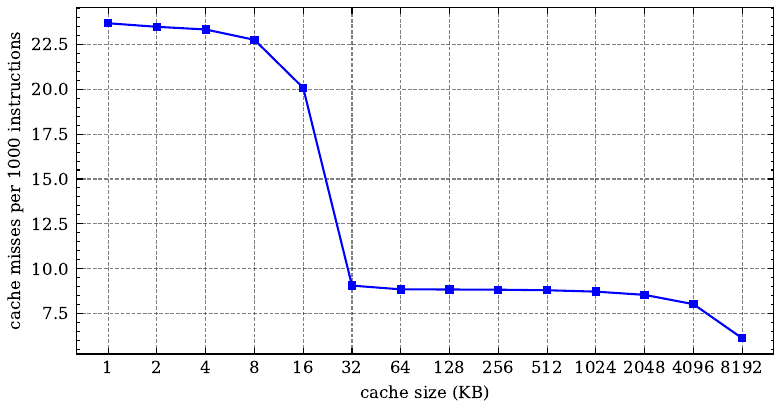}
\caption{stats spc large.log}
\end{figure}

\begin{figure}[t]
\centering
\includegraphics[width=0.9\columnwidth]{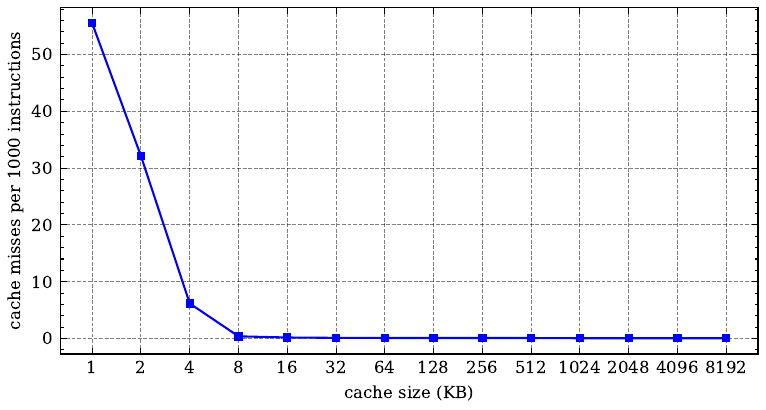}
\caption{stats svm sim fast.log}
\end{figure}

\begin{figure}[t]
\centering
\includegraphics[width=0.9\columnwidth]{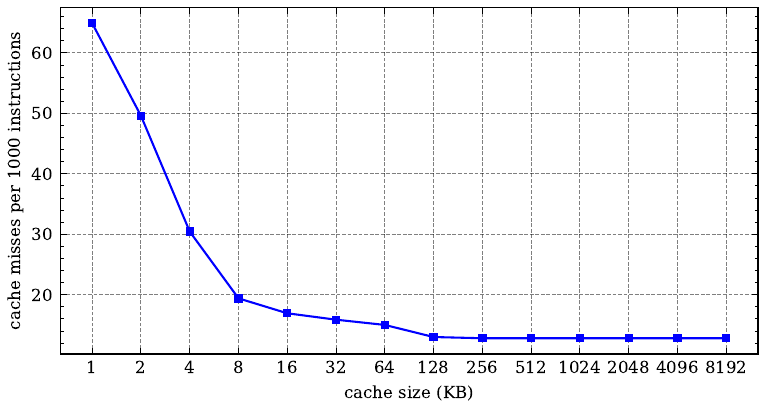}
\caption{stats spc small.log}
\end{figure}

\begin{figure}[t]
\centering
\includegraphics[width=0.9\columnwidth]{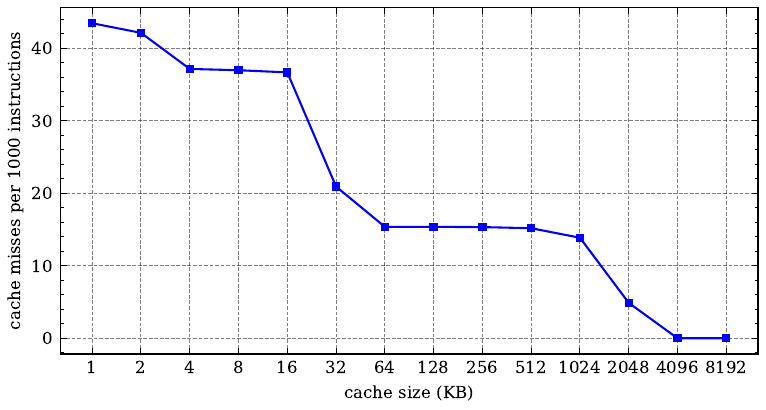}
\caption{stats rbm large.log}
\end{figure}

\begin{figure}[t]
\centering
\includegraphics[width=0.9\columnwidth]{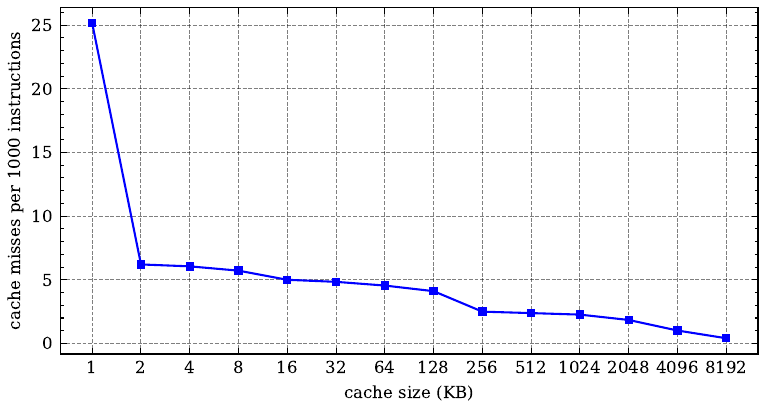}
\caption{stats tracking vga.log}
\end{figure}

\begin{figure}[t]
\centering
\includegraphics[width=0.9\columnwidth]{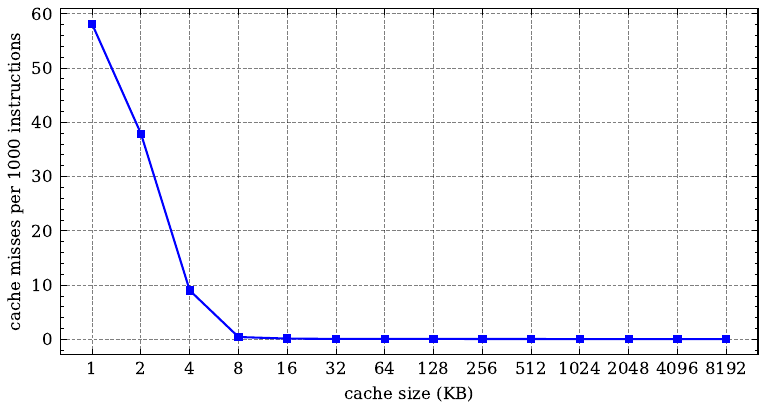}
\caption{stats svm sim.log}
\end{figure}

\begin{figure}[t]
\centering
\includegraphics[width=0.9\columnwidth]{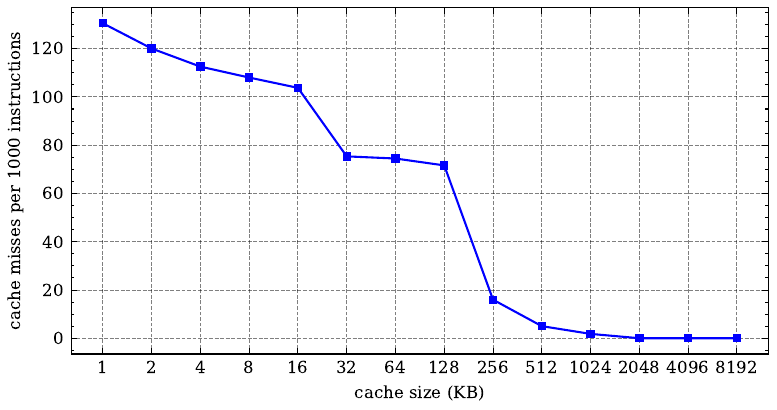}
\caption{stats pca small.log}
\end{figure}

\begin{figure}[t]
\centering
\includegraphics[width=0.9\columnwidth]{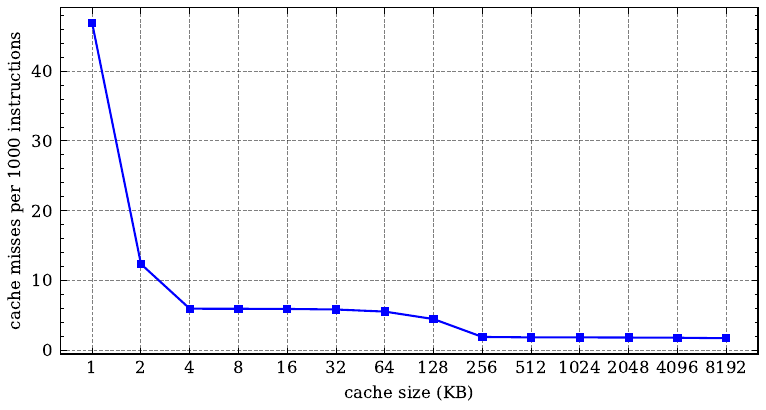}
\caption{stats sift fullhd.log}
\end{figure}

\begin{figure}[t]
\centering
\includegraphics[width=0.9\columnwidth]{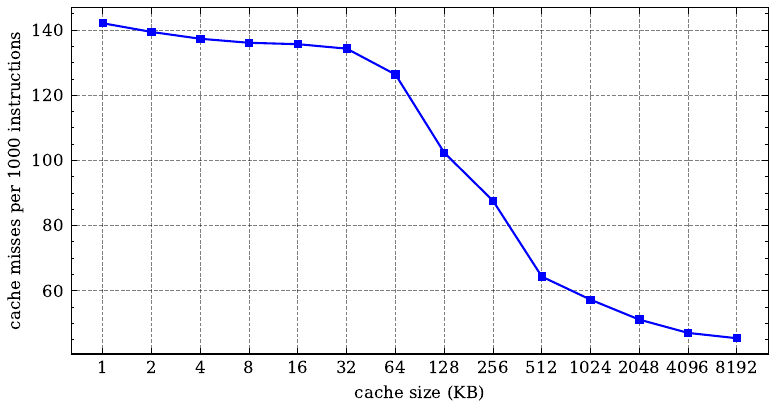}
\caption{stats pca large.log}
\end{figure}

\begin{figure}[t]
\centering
\includegraphics[width=0.9\columnwidth]{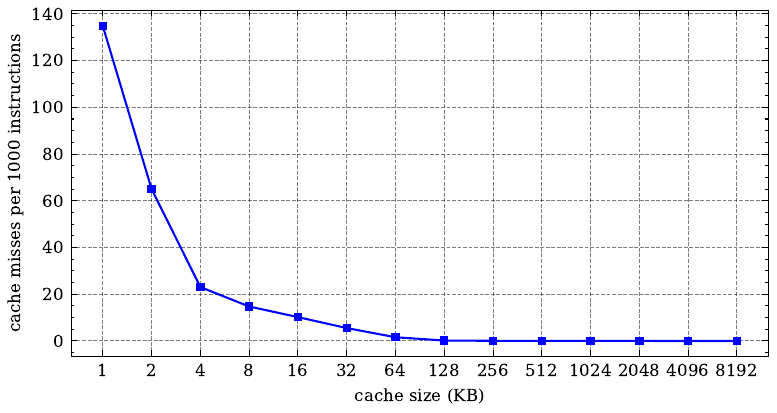}
\caption{stats lda medium.log}
\end{figure}

\begin{figure}[t]
\centering
\includegraphics[width=0.9\columnwidth]{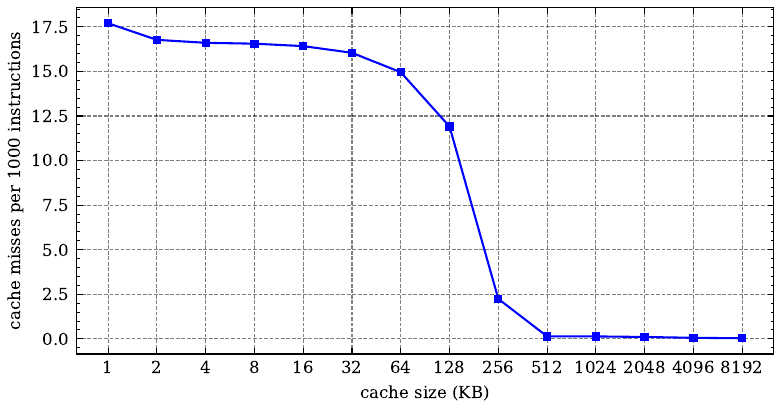}
\caption{stats stitch vga.log}
\end{figure}

\begin{figure}[t]
\centering
\includegraphics[width=0.9\columnwidth]{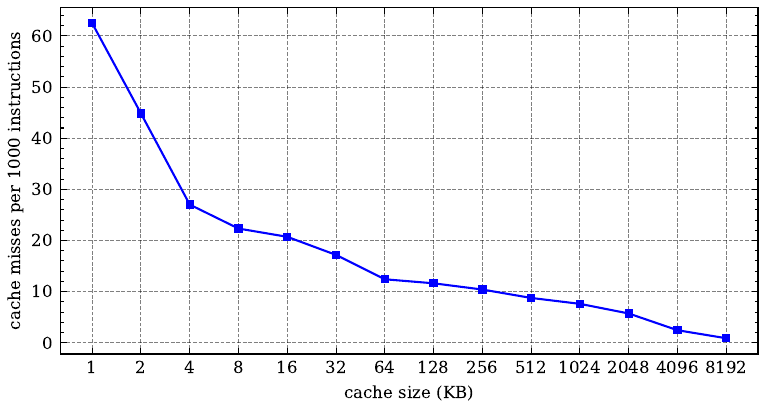}
\caption{stats sphinx medium.log}
\end{figure}

\begin{figure}[t]
\centering
\includegraphics[width=0.9\columnwidth]{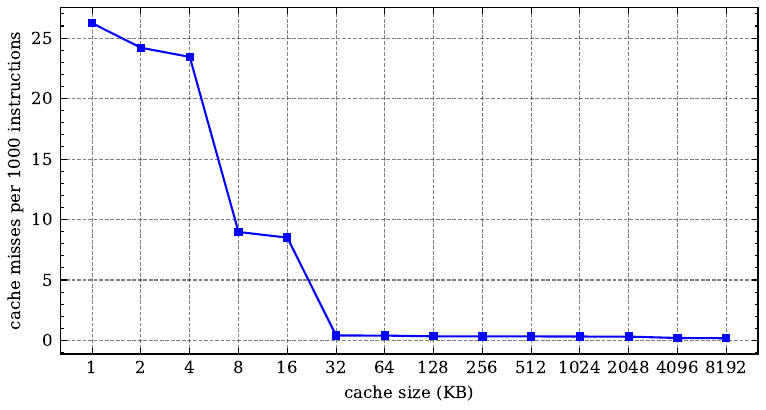}
\caption{stats me large.log}
\end{figure}

\begin{figure}[t]
\centering
\includegraphics[width=0.9\columnwidth]{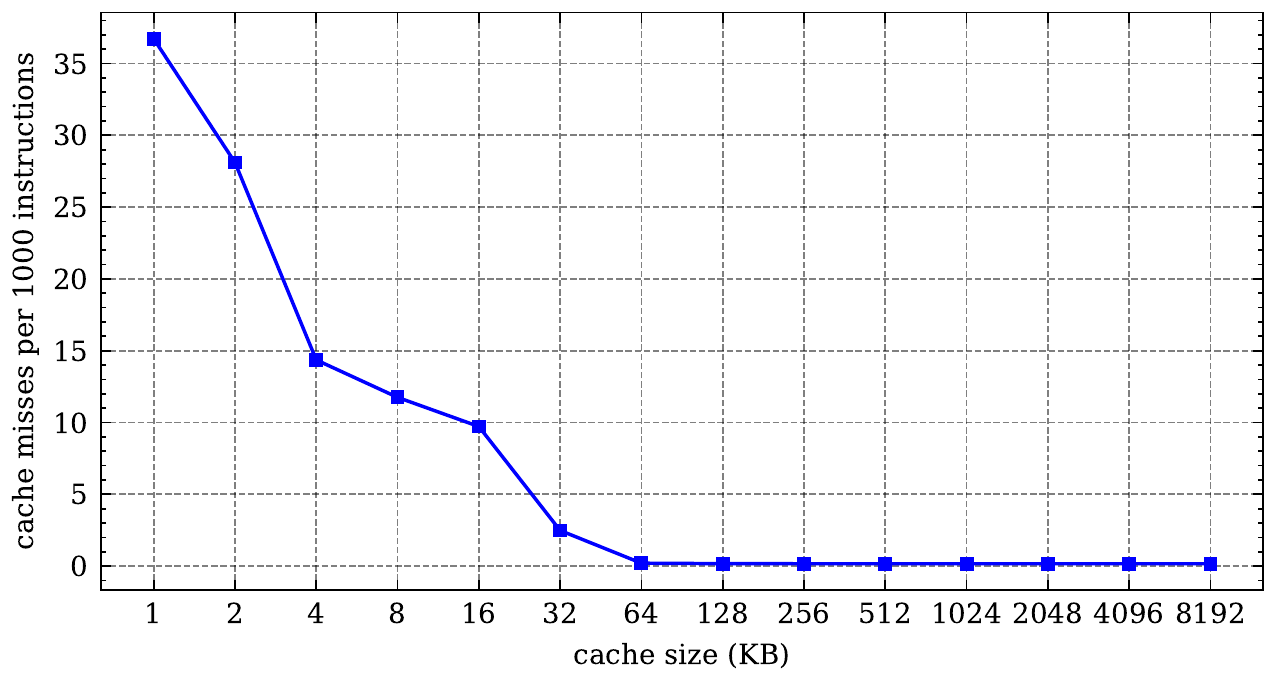}
\caption{stats rbm small.log}
\end{figure}

\begin{figure}[t]
\centering
\includegraphics[width=0.9\columnwidth]{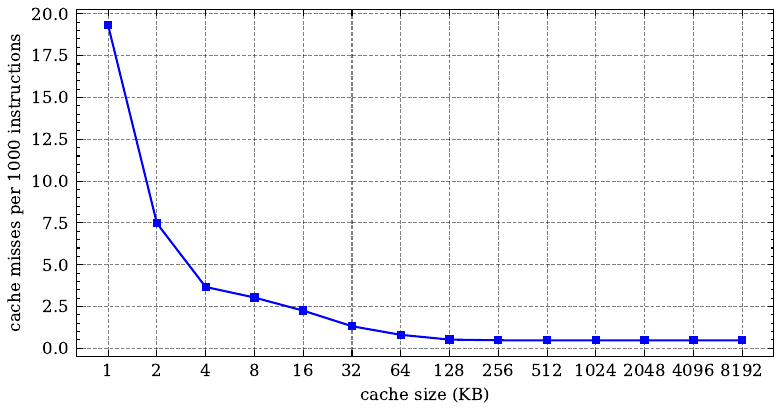}
\caption{stats tracking sim.log}
\end{figure}